\documentclass[12pt]{article}

\usepackage[height=8.5in,width=6.4in]{geometry}

\usepackage{xparse}
\usepackage{amssymb, amsmath}
\usepackage{xcolor}
\usepackage{tikz}
\usepackage{cite}
\usepackage[linktocpage]{hyperref}
\usepackage{mathtools}

\DeclareFontShape{OT1}{cmr}{mx}{n}%
    {<->cmr10}{}

\renewcommand{\tilde}{\widetilde}

\newcommand{\SU}{\text{SU}}
\newcommand{\U}{\text{U}}

\numberwithin{equation}{section}
\def\Nequals#1{$\mathcal{N}{=}\,#1$}

\allowdisplaybreaks

\begin{document}

\begin{titlepage}

\begin{flushright} 
\end{flushright}

\vskip 1.5cm

\begin{center}

{\LARGE\fontseries{mx}\selectfont
Argyres-Douglas theories and Liouville \\ Irregular
 States\par
}

\vskip 1.5cm
{\large
 Takahiro Nishinaka$^{\diamondsuit,1}$ and Takahiro
 Uetoko$^{\clubsuit,1}$
}

\vskip .7cm

{\it
 $^1$
Department of Physical Sciences, College of Science and Engineering\\
 Ritsumeikan University, Shiga 525-8577, Japan
}

\end{center}

\vskip1.5cm

\begin{abstract}
We study irregular states of rank-two and three in Liouville theory, based on an ansatz proposed by
 D.~Gaiotto and J.~Teschner. Using these irregular states, we evaluate
 asymptotic expansions of irregular conformal blocks corresponding to the partition functions
 of $(A_1,A_3)$ and $(A_1,D_4)$ Argyres-Douglas theories for general
 $\Omega$-background parameters.
In the limit of
 vanishing Liouville charge, our result reproduces strong coupling expansions of the partition
 functions recently obtained via
 the Painlev\'e/gauge correspondence.
This suggests that 
the irregular conformal block for one irregular
singularity of rank $3$ on sphere is also related to Painlev\'e II. We also
 find that our partition functions 
are invariant under the action of the
 Weyl group of flavor symmetries once four and two-dimensional
 parameters are correctly identified. We finally propose a
 generalization of this parameter identification to general irregular
 states of integer rank.

\end{abstract}

\renewcommand{\thefootnote}{\fnsymbol{footnote}}
\footnotetext[0]{$^{\diamondsuit}$nishinak@fc.ritsumei.ac.jp, $^{\clubsuit}$rp0019fr@ed.ritsumei.ac.jp}
\renewcommand{\thefootnote}{\arabic{footnote}}

\end{titlepage}

\tableofcontents


\section{Introduction}

Argyres-Douglas (AD) theories are an important series of strongly coupled \Nequals2 superconformal
field theories (SCFTs) in four dimensions \cite{Argyres:1995jj,
Argyres:1995xn, Eguchi:1996vu}. 
Although they are often obtained
as IR SCFTs of RG-flows from  \Nequals2 gauge
theories, their
physics is still to be understood. One
reason for this is that, since the $U(1)_R$ symmetry of the AD theories is
accidental in the RG-flows, it is hard to read off the IR physics 
from its UV 
description.\footnote{For recent developments in
the study of \Nequals1 Lagrangian theories that flow into the AD
theories, see \cite{Maruyoshi:2016tqk, Maruyoshi:2016aim,
Agarwal:2016pjo, Maruyoshi:2018nod, Agarwal:2017roi, Benvenuti:2017kud,
Benvenuti:2017bpg, Giacomelli:2017ckh, Giacomelli:2018ziv, Carta:2018qke}.} The accidental $U(1)_R$ symmetry particularly makes
it difficult to compute the partition functions of the AD theories via
supersymmetric localization. Therefore, until recently, the partition functions of AD
theories were only partially computed for massless cases via the holomorphic
anomaly equations \cite{Huang:2009md}.

However, there has recently been remarkable progress in
the study of the partition functions of AD theories. 
In particular, the authors of \cite{Bonelli:2016qwg} have evaluated strong
coupling expansions of the Nekrasov partition functions of the
$(A_1,A_2),\, (A_1,A_3)$ and $(A_1,D_4)$ AD theories via a remarkable
relation to the tau-functions of Painlev\'e equations, inspired by
\cite{Gamayun:2012ma} and subsequent works \cite{Gamayun:2013auu, Iorgov:2014vla,
Bershtein:2014yia, Gavrylenko:2016zlf}.\footnote{
Another important progress in the study of the partition functions of AD
theories has been made through the E-string theory on $T^2$ \cite{Sakai:2016jdi} and
quantum periods in the
Nekrasov-Shatashvili limit \cite{Ito:2018hwp, Ito:2019twh}.  
}
After these works, the author of \cite{Nagoya:2015cja, Nagoya:2016mlj, Nagoya:2018}
has nicely shown that these strong coupling expansions can be
reproduced from irregular conformal blocks of Virasoro algebra.\footnote{See also \cite{Lisovyy:2018} for closely related
results related to asymptotically free gauge theories.} The emergence of the
two-dimensional conformal field theory is naturally understood via the AGT correspondence \cite{Alday:2009aq, Gaiotto:2009ma} and
its generalization \cite{Bonelli:2011aa ,Gaiotto:2012sf} that relate the Nekrasov
partition functions of 4d \Nequals2 theories to conformal blocks of
two-dimensional Liouville theory. Indeed, an attempt to compute the partition function
of AD theories as Liouville irregular conformal blocks was
partially studied in \cite{Bonelli:2011aa,
Gaiotto:2012sf}.\footnote{For more works on Liouville irregular states,
see \cite{Gaiotto:2009ma, Kanno:2013vi, Matsuo:2014rba, Polyakov:2016noy, Polyakov:2016bjh, Rim:2016msx}.}
Finally, nice matrix model descriptions of this relation between AD
theories and Painlev\'e equations have
been studied in \cite{Grassi:2018spf,
Itoyama:2018wbh, Itoyama:2018gnh}.\footnote{Matrix model
descriptions of the partition functions of 4d $\mathcal{N}=2$ theories
 in the context of the AGT correspondence were studied in various
 papers. See for example \cite{Itoyama:2009sc, Eguchi:2009gf,
 Schiappa:2009cc, Mironov:2009ib, Fujita:2009gf, Itoyama:2010ki,
 Mironov:2010ym, Morozov:2010cq, Eguchi:2010rf, Itoyama:2010na,
 Maruyoshi:2010pw, Bonelli:2010gk, Itoyama:2011mr, Nishinaka:2011aa,
 Galakhov:2012gw, Bourgine:2012gy, Nishinaka:2012kn, Rim:2012tf,
 Choi:2013caa, Itoyama:2013mca, Itoyama:2014pca, Choi:2014qha, Rim:2015tsa, Choi:2015gda, Rim:2015aha,
 Choi:2015idw, Polyakov:2016eir}.}

In this paper, we extend the irregular conformal block approach to
various directions, focusing on the $(A_1,A_3)$ and $(A_1,D_4)$ AD
theories.  First, we compute the irregular conformal blocks for these two
AD theories
using an ansatz proposed in \cite{Gaiotto:2012sf}. This ansatz is based
on the construction of an irregular state in a colliding limit of regular vertex operators, and implies
that a rank-$n$ irregular state is expanded in terms of ``generalized
descendants'' of a rank-$(n-1)$ irregular state. Here, generalized
decendants are descendants in the sense of Virasoro generators and
derivatives with respect to parameters characterizing irregular
states.
Since this construction of irregular conformal blocks is different from
the one used in \cite{Nagoya:2015cja, Nagoya:2016mlj,
Nagoya:2018}, 
it is non-trivial whether it leads to a consistent result.
In this paper, we explicitly show that the ansatz mentioned above gives
 a consistent result, and moreover we demonstrate that the asymptotic expansion
of the ansatz proposed in \cite{Gaiotto:2012sf} precisely corresponds to the strong coupling
expansion of partition functions 
obtained in \cite{Bonelli:2016qwg}.

Second, we reproduce
the partition function of the $(A_1,A_3)$ theory from the rank-three
irregular state of the Liouville theory. This partition function was reproduced in \cite{Nagoya:2018} as the
irregular conformal block with an irregular singular point of
rank $\frac{3}{2}$ and a
regular singular point. On the other hand, from the class $\mathcal{S}$
construction of the $(A_1,A_3)$ theory, the same partition
function is expected to be obtained as the one-point function of an irregular vertex
operator of rank 3. We explicitly show that this is indeed
the case.

Finally, we evaluate these irregular conformal blocks
with non-vanishing Liouville charge
$Q$ turned on.\footnote{Our convention for the Liouville charge is such that the
Virasoro central charge is given by $c=1+6Q^2$.} Since the Painlev\'e tau-functions are related to
the case of $Q=0$, earlier works focused on the vanishing
Liouville charge. On the other hand, in relation to the 4d physics, it is important
to evaluate the irregular conformal blocks for non-vanishing $Q$. Indeed, according to
the AGT correspondence, the Liouville charge $Q$ corresponds to
$(\epsilon_1+\epsilon_2)/\sqrt{\epsilon_1\epsilon_2}$ in four dimensions, where $\epsilon_1$ and
$\epsilon_2$ are two $\Omega$-background parameters
\cite{Nekrasov:2002qd}. Therefore, the non-vanishing $Q$ leads to the partition
function of AD theories with the most
general $\Omega$-background turned on. We particularly show that the
four-dimensional flavor symmetry implies that there exists an $\mathcal{O}(Q)$ correction to the relation between 4d and
2d parameters. We interpret this $\mathcal{O}(Q)$ correction in terms
of the colliding limit of regular vertex operators, which leads us to a
conjecture for a general irregular state of integer rank.

The organization of the paper is the following. In
Sec.~\ref{sec:review}, we briefly review the class
$\mathcal{S}$ construction of AD theories and the generalized AGT correspondence. In
Sec.~\ref{sec:partition}, we evaluate the Liouville irregular states of
rank-two and three, using the ansatz proposed in
\cite{Gaiotto:2012sf}. We then use these irregular states to compute
the irregular conformal blocks corresponding to the partition functions
of $(A_1,D_4)$ and $(A_1,A_3)$ theories for a general $\Omega$-background
parameters $\epsilon_1$ and $\epsilon_2$. In Sec.~\ref{sec:flavor}, we
show that the partition functions of the $(A_1,D_4)$ and $(A_1,A_3)$
theories evaluated in Sec.~\ref{sec:partition} are invariant under
the action of the Weyl group of the enhanced flavor symmetry, once
parameters in four and two dimensions are appropriate identified. 
We also conjecture a general identification of the Liouville momentum of
irregular states with a 4d mass parameter.
In Sec~\ref{sec:conclusion}, we conclude with a comment 
on the symmetry of the partition functions under $(\epsilon_1,\epsilon_2)
\to (-\epsilon_1,-\epsilon_2)$. 

\section{AD theories of class $\mathcal{S}$ and AGT correspondence}
\label{sec:review}

In this section, we briefly review the $(A_1,A_3)$ and
$(A_1,D_4)$ theories as well as the generalized AGT correspondence that
relates the partition functions of AD theories to irregular
conformal blocks of Liouville theory in two dimensions.

\subsection{AD theories of class $\mathcal{S}$}

The AD theories are a series of strongly coupled 4d \Nequals2 SCFTs with
Coulomb branch operators of fractional dimensions. While the simplest AD theories were originally discovered as IR SCFTs at
special points on the Coulomb branch of asymptotically free gauge
theories \cite{Argyres:1995jj,
Argyres:1995xn, Eguchi:1996vu}, their infinite generalizations were
constructed by compactifying 6d $(2,0)$ SCFTs with gauge algebra
$\mathfrak{g}$ on a punctured Riemann surface $\mathcal{C}$
\cite{Bonelli:2011aa, Xie:2012hs}.\footnote{See also \cite{Cecotti:2010fi} for the
construction of AD theories via type II string theory on Calabi-Yau
singularities. Indeed, the AD theories we discuss in
this paper, $(A_1,A_3)$ and $(A_1,D_4)$, were first constructed in this
reference.} 
In this paper, we focus on  $\mathfrak{g}=A_1$. Then the resulting AD
theories are $(A_1,A_N)$ and $(A_1,D_N)$ theories, depending on the
number and types of the punctures on $\mathcal{C}$. In the rest of this
sub-section, we denote by
$\mathcal{T}_{\mathcal{C}}$ the 4d \Nequals2 theory obtained by
compactifying 6d $(2,0)\,$ $A_1$ theory on $\mathcal{C}$.

As mentioned above, the 4d theory $\mathcal{T}_{\mathcal{C}}$ depends on
the number and types of the punctures on $\mathcal{C}$. Possible punctures on
$\mathcal{C}$ are classified into two types; regular and
irregular punctures. To describe these two types,
let us briefly review how the Coulomb branch moduli space of the 4d
theory $\mathcal{T}_{\mathcal{C}}$ is related to the Hitchin system on
$\mathcal{C}$. Since we focus on the case of $\mathfrak{g}=A_1$, we
consider the $A_1$ Hitchin system on $\mathcal{C}$, which involves an
$\SU(2)$-connection $A$ on $\mathcal{C}$ and an $\mathfrak{su}(2)$-valued 1-form $\varphi dz +
\bar{\varphi}d\bar z$ on
$\mathcal{C}$. The pair $(A,\varphi)$ is required to satisfy a set of differential equations called
``Hitchin's equations.'' Since the
Hitchin's equations are differential equations, we need to specify the
boundary conditions of $A$ and $\varphi$ at each
puncture on $\mathcal{C}$. We call a puncture a ``regular puncture'' if
$\varphi$ has a simple pole there,
 while a puncture
is called an ``irregular puncture'' if the singularity of $\varphi$ 
 is not a simple pole. Suppose that an irregular puncture is located at
 $z=0$. When $\varphi$ 
behaves as $\varphi \sim 1/z^{n+1}$ 
around it, we call $n$ the ``rank'' of the irregular
puncture. While the rank could be an integer or a half-integer for $\mathfrak{g}=A_1$, we focus
on integer ranks in this paper. The Hitchin moduli space is then defined as the space of
solutions to Hitchin's equations (modulo gauge transformations) with
the boundary conditions at punctures kept fixed. The spectral curve of the Hitchin system is
defined by
\begin{align}
 \det (xdz - \varphi) =0~.
\label{eq:curve}
\end{align}
A beautiful observation of \cite{Witten:1997sc, Gaiotto:2009we, Gaiotto:2009hg} is that this spectral curve is identified with the Seiberg-Witten (SW) curve of the
corresponding 4d \Nequals2 theory $\mathcal{T}_{\mathcal{C}}$.\footnote{The Seiberg-Witten 1-form is
identified with $xdz$.} 
Note that this SW curve, and therefore the
4d
theory $\mathcal{T}_{\mathcal{C}}$, depend on the number and types of
the punctures on $\mathcal{C}$.
In particular, the flavor symmetry of $\mathcal{T}_{\mathcal{C}}$ is
naturally encoded in the punctures. Indeed,
each regular puncture gives rise to a flavor
$\SU(2)$ symmetry, while
each irregular puncture of integer rank gives rise to a flavor $\mathrm{U}(1)$ symmetry
in $\mathcal{T}_{\mathcal{C}}$.\footnote{Irregular punctures of
non-integer rank do not give rise to a flavor symmetry. We will not
use them in this paper.}
As discussed in \cite{Gaiotto:2009we, Gaiotto:2009hg}, when
$\mathcal{C}$ has no irregular puncture, $\mathcal{T}_{\mathcal{C}}$ is
always an \Nequals2 SCFT.\footnote{To be more precise, $\mathcal{C}$
needs at least three regular punctures for $\mathcal{T}_{\mathcal{C}}$
to be a non-trivial \Nequals2 SCFT.} On the other hand, when
$\mathcal{C}$ has an irregular puncture, the $\mathcal{N}=2$
supersymmetric field
theory $\mathcal{T}_{\mathcal{C}}$ is
conformal if and only if $\mathcal{C}$ is of genus zero, has one or
zero regular puncture, and does not have more than one irregular puncture \cite{Bonelli:2011aa, Xie:2012hs}.

Let us now describe the $(A_1, A_{N})$ and
$(A_1,D_N)$ theories. The $(A_1, A_{2n-3})$ theory is a 4d \Nequals2 SCFT associated with
$\mathcal{C}$ being a sphere with an irregular puncture of rank $n$. On
the other hand, the $(A_1, D_{2n})$ theory is 4d \Nequals2 SCFT associated with $\mathcal{C}$
which is a sphere with a regular puncture and an irregular puncture
of rank $n$. These two theories are AD theories since they have
fractional-dimensional Coulomb branch operators.
 In this paper, we focus on two AD theories in these series;
$(A_1,A_3)$ and $(A_1,D_4)$. These two AD theories are special in the
sense of flavor symmetry enhancement. Indeed, according to the relation
between the flavor symmetry and the punctures on $\mathcal{C}$, the manifest flavor
symmetries of $(A_1,A_{2n-3})$ and $(A_1,D_{2n})$ for
an integer $n$ are
$\U(1)$ and $\SU(2)\times \U(1)$, respectively. However, for $(A_1,A_3)$ and $(A_1,D_4)$
theories, these manifest flavor symmetries get enhanced to $\SU(2)$
and $\SU(3)$, respectively. This flavor symmetry enhancement will be important
in our discussions in Sec.~\ref{sec:flavor}.

For later use, we here write down the SW curves of these
two theories. Up to a change of variables, the SW curve \eqref{eq:curve}
for the $(A_1,A_3)$ theory is
written as\footnote{We here take the holomorphic coordinate $z$ on
$\mathcal{C}$ so that the irregular puncture is located at $z=0$. In the literature, the SW curve of the $(A_1,A_3)$
theory is usually written as
\begin{align}
 x^2 = z^4 + cz^2 + mz + u~,
\end{align}
which is related to \eqref{eq:curve-A3} by the change of coordinates $(x,z)
\to (-z^2x,1/z)$. Note that this coordinate change preserves the SW 1-form $xdz$.
}
\begin{align}
 x^2 = \frac{u}{z^4}+ \frac{m}{z^5} + \frac{c}{z^6}+\frac{1}{z^8}~,
\label{eq:curve-A3}
\end{align}
where $c,m$ and $u$ are coefficients respectively interpreted as a relevant coupling of dimension
$2/3$, a mass parameter, and  the vacuum expectation value (vev) of a
Coulomb branch operator of dimension $4/3$. On the other hand, for the
$(A_1,D_4)$ theory, the SW curve is written as\footnote{
Again, in the literature, this curve is usually written as
\begin{align}
 x^2 = z^2 + cz + m +  \frac{u}{z} + \frac{M^2}{z^2}~,
\end{align}
which is related to \eqref{eq:curve-D4} by $(x,z)\to (-z^2x,1/z)$.
}
\begin{align}
 x^2 =   \frac{M^2}{z^2} + \frac{u}{z^3} + \frac{m}{z^4}  + \frac{c}{z^5} +\frac{1}{z^6}~,
\label{eq:curve-D4}
\end{align}
where $c$ and $u$ are respectively a relevant coupling of dimension
$1/2$ and the vev of a Coulomb branch operator of dimension $3/2$, while
$m$ and $M$ are two mass parameters.
See \cite{Bonelli:2011aa, Xie:2012hs} for more detail.

\subsection{Irregular states and generalized AGT correspondence}
\label{subsec:AGT}

We here review the generalized AGT correspondence that relates the
partition functions of AD theories to irregular conformal blocks of
two-dimensional Liouville theory. Let
$\mathcal{T}_{\mathcal{C}}$ be a 4d \Nequals2 SCFT associated with a
Riemann surface $\mathcal{C}$ with only regular punctures. Then the AGT correspondence
\cite{Alday:2009aq} states that the Nekrasov
partition functions \cite{Nekrasov:2002qd} of $\mathcal{T}_{\mathcal{C}}$ to the
conformal block of Liouville theory on $\mathcal{C}$. Here, each
regular puncture on $\mathcal{C}$ corresponds to the insertion of a
Virasoro 
primary vertex operator $e^{2\gamma\phi(z)}$ whose Liouville momentum $\gamma$ is fixed by the mass
parameter associated with the regular puncture. Therefore, when
$\mathcal{C}$ has $\ell$ regular punctures, then the partition function of
$\mathcal{T}_{\mathcal{C}}$ is identified with the $\ell$-point conformal
block of Liouville theory, i.e.,
\begin{align}
 Z_{\mathcal{T}_{\mathcal{C}}}^{\epsilon_1,\epsilon_2} =
 \mathcal{F}_{\gamma_1}{}^{\gamma_2}{}_{\gamma_3}\cdots{}^{\gamma_{\ell-1}}{}_{\gamma_\ell}~,
\end{align}
where $\epsilon_1$ and $\epsilon_2$ are the $\Omega$-background parameters
turned on in four dimensions. In this paper, we fix the
scale of the LHS by setting $\epsilon_1\epsilon_2=1$ as in
\cite{Alday:2009aq}. The ratio of the $\Omega$-background parameters is
related to the Liouville charge $Q$ by\footnote{Recall here that our convention for $Q$  is such that the Virasoro central
charge is given by $c=1+6Q^2$.}
\begin{align}
 Q=b+\frac{1}{b}~,\qquad b^2\equiv \frac{\epsilon_1}{\epsilon_2}~.
\label{eq:Omega}
\end{align}
It was also proposed in \cite{Alday:2009aq} that the SW curve of
$\mathcal{T}_{\mathcal{C}}$ is identified as the Liouville correlator
with a stress tensor insertion:
\begin{align}
 x^2 = \frac{\langle \Delta_{\gamma_1}|T(z)
 e^{2\gamma_2 \phi}\cdots e^{2\gamma_{\ell-1}\phi}|\Delta_{\gamma_{\ell}}\rangle
 }{\langle \Delta_{\gamma_1}|e^{2\gamma_2\phi}\cdots e^{2\gamma_{\ell-1}\phi}
| \Delta_{\gamma_\ell}\rangle}~,
\label{eq:curve-from-2d}
\end{align}
where $|\Delta_\gamma\rangle$ is the Virasoro primary state of
holomorphic dimension $\Delta_\gamma \equiv \gamma(Q-\gamma)$. 

A generalization of the AGT correspondence to AD theories has been discussed in
\cite{Bonelli:2011aa, Gaiotto:2012sf},\footnote{See also
\cite{Gaiotto:2009ma} for the first generalization to class
$\mathcal{S}$ theories whose Riemann surface involves irregular
punctures.} in which an irregular puncture
of rank $n$ is conjectured to correspond to a particular linear
combination, $|I^{(n)}\rangle$, of
a Virasoro primary and its descendants.
In this paper, we follow the construction of $|I^{(n)}\rangle$ proposed in \cite{Gaiotto:2012sf}.
Namely, the state $|I^{(n)}\rangle$ is a solution to the
following $2n$ equations involving $(n+1)$ parameters
$c_1,\cdots,c_n$ and $\alpha$:
\begin{align}
 L_{k}|I^{(n)}\rangle &= \left\{
\begin{array}{l}
 \Lambda_k |I^{(n)}\rangle\qquad \text{for}\quad n\leq k\leq 2n
\\[1mm]
\left(\Lambda_k +\sum_{\ell=k+1}^{n-1}(\ell-k)c_\ell\frac{\partial}{\partial c_{\ell-k}}\right)|I^{(n)}\rangle \qquad \text{for}\quad 1\leq k\leq n-1
\\[1mm]
\left(\Delta_\alpha + \sum_{\ell=1}^{n}\ell c_\ell \frac{\partial}{\partial c_\ell}\right)|I^{(n)}\rangle \qquad \text{for} \quad k=0
\\
\end{array}
\right.~,
\label{eq:irregular}
\end{align}
where $L_k$ are the generators of Virasoro algebra at central charge
 $c=1+6Q^2$,
and
\begin{align}
 \Lambda_k\equiv \left\{
\begin{array}{l}
-\sum_{\ell=k-n}^{n}c_\ell c_{k-\ell} \qquad \text{for}\quad n+1\leq k \leq 2n
\\[2mm]
 -\sum_{\ell=1}^{n-1}c_\ell c_{n-\ell} + \big((k+1)Q-2\alpha\big)c_k\qquad \text{for}\quad 1\leq k\leq n
\\
\end{array}
\right.~.
\end{align}
We call $|I^{(n)}\rangle$ satisfying the above equations  an ``irregular
state'' of rank $n$, or simply a rank-$n$ state. Note that the rank-zero
state $|I^{(0)}\rangle$ is naturally identified with a regular primary
state of Virasoro algebra.

Note here that  $|I^{(n)}\rangle$ is a simultaneous eigenstate of
$L_n,\cdots,L_{2n-1}$ and $L_{2n}$ with eigenvalues fixed by $c_1,\cdots,c_{n}$ and $\alpha$.\footnote{To be precise,
eigenstates of $L_n,\cdots,L_{2n}$ were considered in
\cite{Gaiotto:2009ma, Gaiotto:2012sf}, while \cite{Bonelli:2011aa}
studied eigenstates of $L_1$ and $L_{2n}$. The relation between these two were
discussed in appendix A of \cite{Kanno:2013vi}. These two descriptions
are both useful in understanding these special states.}
This particularly implies that $L_{k}|I^{(n)}\rangle = 0$ for all
$k>2n$. 
 We particularly call $\alpha$ the ``Liouville
momentum'' of $|I^{(n)}\rangle$, which is natural in the following
sense. In \cite{Gaiotto:2012sf}, the rank-$n$ state $|I^{(n)}\rangle$ was
 constructed in a colliding limit of $n$ regular vertex
operators $e^{2\gamma \phi(z)}$ to a regular primary state
$|\Delta_{\gamma_0}\rangle$.\footnote{See Sec.~2.2 in particular. See
also Sec.~2 of \cite{Nishinaka:2012kn} for a review.} Namely, $|I^{(n)}\rangle$ is obtained from
\begin{align}
 e^{2\gamma_{n} \phi(z_n)} e^{2\gamma_{n-1}\phi(z_{n-1})}\cdots
 e^{2\gamma_1\phi(z_1)}|\Delta_{\gamma_{0}}\rangle~,
\label{eq:colliding0}
\end{align}
in the limit $z_1,\cdots,z_n\to 0$. For this limit to make sense, we
need to take $\gamma_0,\cdots,\gamma_n
\to\infty$ simultaneously with 
$\sum_{i=0}^n \gamma_i$ 
 kept fixed.
This sum of $\gamma_i$ is identified with $\alpha$ in \eqref{eq:irregular},
i.e., $\alpha\equiv \sum_{i=0}^n\gamma_i$. Therefore, $\alpha$ characterizing
$|I^{(n)}\rangle$ is the sum of the
Liouville momenta of regular vertex operators involved in the
colliding limit. This point will be
important in Sec.~\ref{sec:flavor}.

On the other hand, $|I^{(n)}\rangle$ is {\it not} an eigenstate of
$L_0,\cdots,L_{n-2}$ or $L_{n-1}$. Indeed, the actions of
$L_0,\cdots,L_{n-1}$ involve derivatives with respect to $c_i$. This
means that
\eqref{eq:irregular} are differential equations, and therefore $|I^{(n)}\rangle$ is
not uniquely fixed by $c_1,\cdots,c_{n}$ and $\alpha$. To fix
$|I^{(n)}\rangle$ completely, we need to specify its ``boundary condition'' or
``asymptotic behavior.'' In
this paper, we assume the following
asymptotic behavior of $|I^{(n)}\rangle$ proposed in
\cite{Gaiotto:2012sf}; in the small $c_n$ limit, $|I^{(n)}\rangle$
behaves as
\begin{align}
 |I^{(n)}(\alpha,\pmb{c})\rangle = f(\alpha,\pmb{c},\beta_{n-1})\left(|I^{(n-1)}(\beta_{n-1},\widetilde{\pmb{c}})\rangle + \sum_{k=1}^{\infty}
 (c_n)^{k} |I^{(n-1)}_{k}\rangle\right)~,
\label{eq:general-ansatz}
\end{align}
where $\pmb{c}\equiv (c_1,\cdots,c_n)$, $\widetilde{\pmb{c}}\equiv (
c_1,\cdots,c_{n-1})$, and $\beta_{n-1}$ is a free parameter
characterizing the asymptotic behavior.\footnote{See in
particular Sec.~3.2 and Appendix B.2 of the reference
\cite{Gaiotto:2012sf} for the first discussion of this ansatz.
We also mention here that $|I^{(n-1)}_k\rangle$ in
\eqref{eq:general-ansatz} was denoted as
$|I^{(n-1)}_{nk}\rangle$ in \cite{Gaiotto:2012sf}. We use
$|I^{(n)}_k\rangle$ instead of $|I^{(n)}_{nk}\rangle$ to reduce clutter.
} On the RHS of \eqref{eq:general-ansatz},
 the leading term $|I^{(n-1)}(\beta_{n-1},\widetilde{\pmb{c}})\rangle$ is the rank-$(n-1)$
irregular state satisfying \eqref{eq:irregular} with $n$ and $\alpha$
replaced by $(n-1)$ and $\beta_{n-1}$, respectively. Note that $\beta_{n-1}$ is the Liouville momentum of this leading term. On the other hand, the
remaining terms
$|I^{(n-1)}_{k}\rangle$ are {\it generalized descendants} of
$|I^{(n-1)}(\beta_{n-1},\widetilde{\pmb{c}})\rangle$. Here, generalized descendants are descendants in
the sense of Virasoro generators $L_{k<0}$ and the derivatives with respect to
$c_1,\cdots,c_{n-1}$.\footnote{These derivatives make sense since
 $|I^{(n-1)}(\beta_{n-1},\widetilde{\pmb{c}})\rangle$ depends
on $c_1,\cdots, c_{n-1}$.}
The prefactor $f(\alpha,\pmb{c},\beta_{n-1})$ in
\eqref{eq:general-ansatz} is a function necessary for the
ansatz \eqref{eq:general-ansatz} to satisfy the differential equations
\eqref{eq:irregular} order by order in $c_n$.\footnote{For concrete expressions for
$f(\alpha,\pmb{c};\beta)$ for $n=2$ and $3$, see \eqref{eq:expansion-D4}
and \eqref{eq:expansion-A3} in Sec.~\ref{sec:partition}.} 

Note here that $|I^{(n-1)}(\beta_{n-1},\widetilde{\pmb{c}})\rangle$ on
the RHS of \eqref{eq:general-ansatz} can further be expanded in terms of
rank-$(n-2)$ state and its generalized descendants. In this expansion,
the Liouville momentum of the rank-$(n-2)$ state is again a free
parameter, which we denote by $\beta_{n-2}$. Continuing this
expansion until we reach the rank-zero state (i.e., regular primary
state), we see that there are $n$ free parameters $\beta_{0},\cdots,\beta_{n-1}$ characterizing the original rank-$n$
state $|I^{(n)}\rangle$.
This means that the
rank-$n$ state $|I^{(n)}\rangle$ depends not only on $\alpha$
and $\pmb{c}=(c_1,\cdots,c_n)$ but also on $\pmb{\beta}\equiv
(\beta_0,\cdots,\beta_{n-1})$. In particular, $\pmb{\beta}=(\beta_0,\cdots,\beta_{n-1})$ are parameters
completely fixing the asymptotic behavior of $|I^{(n)}\rangle$.
Therefore, in the rest of this paper, we write
$|I^{(n)}(\alpha,\pmb{c};\pmb{\beta})\rangle$ instead of $|I^{(n)}(\alpha,\pmb{c})\rangle$ to denote the
Liouville irregular state of rank $n$. This asymptotic expansion of the rank-$n$
irregular state is proposed in \cite{Gaiotto:2012sf} based on the
colliding limit of regular vertex operators explained around \eqref{eq:colliding0}.

Given the above correspondence between irregular punctures and Liouville
 irregular states, the partition functions of the $(A_1,A_{2n-3})$ and
 $(A_1,D_{2n})$ theories are expected to be identified with Liouville
 correlators involving an irregular state. In particular, since  $\mathcal{C}$ for the
 $(A_1,A_{2n-3})$ theory is  a sphere with one irregular
 puncture of rank $n$, its partition function is expected to be written as
\begin{align}
 \mathcal{Z}^{\epsilon_1,\epsilon_2}_{(A_1,A_{2n-3})}= \langle 0 | I^{(n)}(\alpha,\pmb{c};\pmb{\beta})\rangle~,
\label{eq:AGT-AA}
\end{align}
where $|0\rangle$ is the vacuum. On the other hand, $\mathcal{C}$
for the $(A_1,D_{2n})$
theory is a sphere with one regular puncture and one
irregular puncture of rank $n$, and therefore its partition function is expected to be
given by
\begin{align}
  \mathcal{Z}^{\epsilon_1,\epsilon_2}_{(A_1,D_{2n})} = \langle \Delta_\gamma |
 I^{(n)}(\alpha,\pmb{c};\pmb{\beta})\rangle~.
\label{eq:AGT-AD}
\end{align}
In the next section, we explicitly identify the irregular states of rank
2 and 3, and then evaluate the partition functions for $(A_1,A_3)$ and
$(A_1,D_4)$ theories via the relations \eqref{eq:AGT-AA} and \eqref{eq:AGT-AD}.

\section{Partition functions from irregular states}

\label{sec:partition}

In this section, we explicitly compute the Liouville irregular state of
rank $2$ and $3$, and then evaluate the partition functions of the $(A_1,A_3)$
and $(A_1,D_4)$ theories via \eqref{eq:AGT-AA} and \eqref{eq:AGT-AD}.

\subsection{$(A_1,D_4)$ theory from rank-2 irregular state}
\label{sec:A1D4}

Let us first focus on the rank-2 irregular state and the $(A_1,D_4)$ theory.
As reviewed in the previous section, the partition function of the
$(A_1,D_4)$ theory is expected to be given by 
\begin{align}
Z_{(A_1,D_4)}^{\epsilon_1,\epsilon_2} = \langle \Delta_\gamma |
 I^{(2)}(\alpha,\pmb{c};\pmb{\beta})\rangle~,
\end{align}
where $|\Delta_\gamma\rangle$ is a regular primary state of holomorphic
dimension $\Delta_\gamma\equiv \gamma(Q-\gamma)$,
$\pmb{c}$ and $\pmb{\beta}$ are respectively short-hand notations for
$(c_1,c_2)$ and $(\beta_0,\beta_1)$, and
$|I^{(2)}(\alpha,\pmb{c};\pmb{\beta})\rangle$ is the irregular state of rank two satisfying
\begin{align}
L_4|I^{(2)}(\alpha,\pmb{c};\pmb{\beta})\rangle &= -c_2^2|I^{(2)}(\alpha,\pmb{c};\pmb{\beta})\rangle~,
\label{eq:L4-D4}
\\
L_3 |I^{(2)}(\alpha,\pmb{c};\pmb{\beta})\rangle &= -2c_1c_2|I^{(2)}(\alpha,\pmb{c};\pmb{\beta})\rangle~,
\\
L_2 |I^{(2)}(\alpha,\pmb{c};\pmb{\beta})\rangle &= -\left(c_1^2 + c_2(2\alpha-3Q)\right)|I^{(2)}(\alpha,\pmb{c};\pmb{\beta})\rangle~,
\\
L_1 |I^{(2)}(\alpha,\pmb{c};\pmb{\beta})\rangle &= \left(c_2
 \frac{\partial}{\partial
 c_1}-2c_1(\alpha-Q)\right)|I^{(2)}(\alpha,\pmb{c};\pmb{\beta})\rangle~,
\label{eq:L1-D4}
\\
L_0|I^{(2)}(\alpha,\pmb{c};\pmb{\beta})\rangle &= \left(\Delta_\alpha + c_1\frac{\partial}{\partial c_1} + 2c_2\frac{\partial}{\partial c_2}\right)|I^{(2)}(\alpha,\pmb{c};\pmb{\beta})\rangle~.
\label{eq:L0-D4}
\end{align}
To identify the relation between the 4d and 2d parameters, let
us first compute the SW-curve in terms of the irregular state using the
identification \eqref{eq:curve-from-2d}. Since the Virasoro stress tensor
is expanded as 
$T(z) = \sum_{n=-\infty}^{\infty}L_n z^{-n-2}$, the above equations and
$\langle\Delta_\gamma|L_0 = \langle\Delta_\gamma|\Delta_\gamma$ imply that
\begin{align}
\frac{\langle \Delta_\gamma |T(z)| I^{(2)}(\alpha,\pmb{c};\pmb{\beta})\rangle}{\langle \Delta_\gamma| I^{(2)}(\alpha,\pmb{c};\pmb{\beta})\rangle} 
= \frac{\Delta_\gamma}{z^2} + \frac{\sqrt{-ic_2}\,u}{z^3} - \frac{c_1^2 + c_2(2\alpha-3Q)}{z^4} -\frac{2c_1c_2}{z^5} -\frac{c_2^2}{z^6}~,
\end{align}
where
$
u \equiv -\sqrt{-ic_2}\frac{\partial}{\partial c_1}\log \,\langle \Delta_\gamma |
 I^{(2)}(\alpha,\pmb{\beta},\pmb{c})\rangle -\frac{2c_1}{\sqrt{ic_2}}(\alpha-Q)
$. To make a connection to the SW curve \eqref{eq:curve-D4} of the
$(A_1,D_4)$ theory, let us rescale the coordinate as $z \to
\sqrt{-ic_2}\,z$. Since the stress tensor $T$ is of dimension two, this also
rescales $T$ as $T\to -(c_2)^{-2}T$.\footnote{In other words, $T(z)dz$ is
invariant under this transformation.} As a result, we have
\begin{align}
\frac{\langle \Delta_\gamma |T(z)|
 I^{(2)}(\alpha,\pmb{c};\pmb{\beta})\rangle}{\langle \Delta_\gamma|
 I^{(2)}(\alpha,\pmb{c};\pmb{\beta})\rangle} 
= \frac{\Delta_\gamma}{z^2} + \frac{u}{z^3} + \frac{\frac{c_1^2}{c_2} + (2\alpha-3Q)}{z^4} +\sqrt{\frac{ic_1^2}{c_2}}\frac{1}{z^5} +\frac{1}{z^6}~.
\end{align}
The identification \eqref{eq:curve-from-2d} then gives us exactly the
same curve as the SW-curve \eqref{eq:curve-D4} of the $(A_1,D_4)$
theory. From this curve, we can read off how the 2d parameters are
related to the four-dimensional mass parameters, relevant coupling, and
the vev of the Coulomb branch operator. In particular, we see that
$\sqrt{ic_1^2/c_2}$ is identified with the relevant coupling of dimension
$1/2$.

\subsubsection{Ansatz in terms of generalized descendants}

Given the identification of parameters, we now compute the rank-2
irregular state $|I^{(2)}(\alpha,\pmb{c};\pmb{\beta})\rangle$.
To that end, we use the ansatz \eqref{eq:general-ansatz} for the
irregular state reviewed in
Sec.~\ref{sec:review}. For the rank-$2$ irregular state, the ansatz
is written as
\begin{align}
|I^{(2)}(\alpha,\pmb{c};\pmb{\beta})\rangle &=
 c_1^{\nu_1}c_2^{\nu_2}e^{(\alpha-\beta_1)\frac{c_1^2}{c_2}}\bigg(|I^{(1)}(\beta_1,c_1;\beta_0)\rangle
 + \sum_{k=1}^\infty
 (c_2)^k|I^{(1)}_{k}\rangle\bigg)~,
\label{eq:expansion-D4}
\end{align}
where the prefactor $f(\alpha,\pmb{c},\beta_1) = c_1^{\nu_1}c_2^{\nu_2}e^{(\alpha-\beta_1)\frac{c_1^2}{c_2}}$ was identified in \cite{Gaiotto:2012sf}.
The states $|I^{(1)}_{k}\rangle$ are  generalized
descendants of the rank-one irregular state
$|I^{(1)}(\beta_1,c_1;\beta_0)\rangle$.\footnote{As mentioned already, we
here use a slightly different convention for $|I^{(2)}_k\rangle$.
Namely, $|I^{(2)}_k\rangle$ here is identified with $|I^{(2)}_{2k}\rangle$ in
\cite{Gaiotto:2012sf}.} 
As reviewed in Sec.~\ref{sec:review}, a generalized descendant is a descendant of
$|I^{(1)}(\beta_1,c_1;\beta_0)\rangle$ in the sense of $L_{-\ell}$ and
$\partial/\partial c_1$, i.e., a linear combination of states of
the form
\begin{align}
 (L_{-\ell_1})^{p_1}(L_{-\ell_2})^{p_2}\cdots
 (L_{-{\ell_n}})^{p_n}\frac{\partial^m}{\partial c_1{}^m}|I^{(1)}(\beta_1,c_1;\beta_0)\rangle~,
\end{align}
for integers $m,n\geq 0$,\, $p_i>0$, and $\ell_1> \ell_2>\cdots > \ell_n>0$.
The coefficients of this linear combination turn out to
 depend on 
 $c_1$ and $\beta_1$ but are independent of $\beta_0$. As pointed out in
 \cite{Gaiotto:2012sf}, $|I^{(1)}_{k}\rangle$ turns out to be a
 level-$2k$ generalized descendant, where $L_{-k}, \partial^k/\partial c_1^k$ and
 $1/c_1^k$ are regarded as level $k$.
The exponents $\nu_1$ and $\nu_2$ in \eqref{eq:expansion-D4} are fixed as \cite{Gaiotto:2012sf}
\begin{align}
\nu_1 = 2(\alpha-\beta_1)(Q-\beta_1)~,\qquad \nu_2 = \frac{1}{2}(\beta_1-\alpha)(3Q-3\beta_1-\alpha)~,
\end{align}
in order for the ansatz \eqref{eq:expansion-D4} to satisfy
\eqref{eq:L4-D4} -- \eqref{eq:L0-D4}. Note here that the ansatz \eqref{eq:expansion-D4}
particularly implies that in the limit $c_2\to 0$ the irregular state
behaves as $|I^{(2)}(\alpha,\pmb{c};\pmb{\beta})\rangle \to
c_1^{\nu_1}c_2^{\nu_2}e^{(\alpha-\beta_1)\frac{c_1^2}{c_2}}|I^{(1)}(\beta_1,c_1;\beta_0)\rangle$. Therefore
\eqref{eq:expansion-D4} fixes the ``boundary condition'' for
$|I^{(2)}(\alpha,\pmb{c};\pmb{\beta})\rangle$ at $c_2=0$.
The authors of \cite{Gaiotto:2012sf} conjectured that \eqref{eq:L4-D4}
-- \eqref{eq:L0-D4} uniquely determine the generalized descendants
$|I^{(1)}_{k}(\alpha,c_1;\pmb{\beta})\rangle$. 

We now explicitly compute $|I^{(1)}_{k}(\alpha,c_1;\pmb{\beta})\rangle$
by solving \eqref{eq:L4-D4} -- \eqref{eq:L0-D4} order by order of
$c_2$. Since the basic strategy is already discussed in
\cite{Gaiotto:2012sf}, we move the detailed description for the
order-by-order computation to appendix
\ref{app:detail-D4}, and write down the results here:
\begin{align}
|I^{(1)}_0\rangle &= |I^{(1)}(\beta_1,c_1;\beta_0)\rangle~,
\label{eq:descendant1}
\\
|I^{(1)}_1\rangle &= \left(\frac{1}{2c_1}L_{-1} +
 \frac{2\alpha-3\beta_1}{2c_1}\partial_{c_1} +
 \frac{\nu_3}{c_1^2}\right)|I^{(1)}(\beta_1,c_1;\beta_0)\rangle~,
\label{eq:descendant2}
\\
|I^{(1)}_2\rangle &= \bigg(
-\frac{1}{4c_1^2}L_{-2} +\frac{1}{8c_1^2}L_{-1}^2+
 \frac{2\alpha-3\beta_1}{4c_1^2}L_{-1}\partial_{c_1} 
\nonumber\\
&+
 \frac{2(2\alpha-3\beta_1)^2-1}{16c_1^2}\partial_{c_1}^2 + \frac{11\beta_1-8\alpha+Q+4\nu_3}{8c_1^3}L_{-1} 
\nonumber\\
& + \frac{-4\alpha^2+8\alpha(3\beta_1+\nu_3) - \beta_1(23\beta_1+12\nu_3) + Q(-12\alpha+11\beta_1)}{8c_1^3}\partial_{c_1} + \frac{\nu_4}{c_1^4}
\bigg)|I^{(1)}(\beta_1,c_1;\beta_0)\rangle~,
\label{eq:descendant3}
\end{align}
where $\partial_{c_1}\equiv \partial/\partial c_1$, and $\nu_3$ and $\nu_4$ are given by
\begin{align}
\nu_3 &\equiv \frac{1}{2}(3Q+\alpha-3\beta_1)(\alpha-\beta_1)(Q-\beta_1)~,
\\
\nu_4 &\equiv \frac{(Q-\beta_1)(\alpha-\beta_1)}{8}\Big((\alpha -\beta_1 )^3 (Q-\beta_1 )+(\alpha -\beta_1 )^2 \left(4 \beta_1 ^2+6 Q^2-10 \beta_1  Q-2\right)
\nonumber\\
&\qquad \qquad +(\alpha -\beta_1 ) (3 Q-2 \beta_1 ) \left(2 \beta_1 ^2+3 Q^2-5 \beta_1  Q-5\right)-8 \beta_1^2-21 Q^2+26 \beta_1  Q+1
\Big)~.
\end{align}
From the fact that $|I^{(1)}(\beta_1,c_1;\beta_0)\rangle$ is an irregular
state of rank one, we see that \eqref{eq:expansion-D4} with the above
expressions substituted correctly satisfies the equations
\eqref{eq:L4-D4} -- \eqref{eq:L0-D4}.

\subsubsection{Irregular conformal block}

Using the above result for the irregular state of rank two, we here evaluate the $c_2$-expansion of the inner product $\langle
\Delta_\gamma | I^{(2)}(\alpha,\pmb{c};\pmb{\beta})\rangle$. We
particularly show that, in the limit of $Q\to 0$, this $c_2$-expansion is
precisely identical to the strong coupling expansion of the partition function of
the $(A_1,D_4)$ theory obtained in
\cite{Bonelli:2016qwg}. 

To that end, we first evaluate the inner product of \eqref{eq:expansion-D4}
and $|\Delta_\gamma\rangle$, which reduces to computing the inner
product $\langle \Delta_\gamma|I^{(1)}_{k}\rangle$ for all $k$.
Note here that, since
$\langle \Delta_\gamma|L_{-n}=0$ for $n>1$, we only need to keep track
of terms without $L_{-n}$ in \eqref{eq:descendant1} --
\eqref{eq:descendant3}. Moreover, since the rank-one irregular state
satisfies $L_0|I^{(1)}(\beta_1,c_1;\beta_0)\rangle =
\left(\Delta_{\beta_1} +
c_1\partial_{c_1}\right)|I^{(1)}(\beta_1,c_1;\beta_0)\rangle$, the
action of $\partial_{c_1}$ on $|I^{(1)}(\beta_1,c_1;\beta_0)\rangle$ is translated into the
action of $L_0$. Therefore, 
\begin{align}
\langle \Delta_\gamma|
(c_1\partial_{c_1})^k|I^{(1)}(\beta_1,c_1;\beta_0)\rangle =
(\Delta_\gamma-\Delta_{\beta_1})^k\langle \Delta_\gamma|I^{(1)}(\beta_1,c_1;\beta_0)\rangle~.
\end{align}
Using this identity, we see that the inner product of
$|I^{(2)}(\alpha,\pmb{\beta},\pmb{c})\rangle$ and
$|\Delta_\gamma\rangle$ is written as
\begin{align}
\langle \Delta_\gamma|I^{(2)}(\alpha,\pmb{c};\pmb{\beta})\rangle &=
 \langle \Delta_\gamma | I^{(1)}(\beta_1,c_1;\beta_0)\rangle
 c_1^{\nu_1}c_2^{\nu_2}e^{(\alpha-\beta_1)\frac{c_1^2}{c_2}}\sum_{k=0}^\infty
 \left(\frac{c_2}{ic_1^2}\right)^kD_k(\alpha,\beta_1,\gamma)~,
\label{eq:result-D4}
\end{align}
with coefficients $D_k$ depending on $\alpha,\beta_1$ and $\gamma$. We here
show the expressions for the first several coefficients:
\begin{align}
 D_0 &= 1~,
\label{eq:Q-dependent-D4-D0}
\\
D_1 &= \frac{i}{2} \Big(\left(2 \alpha -3 \beta _1\right) \Delta
 _{\gamma }+\left(Q-\beta _1\right) \left(6 \beta _1^2-3 \beta _1 (2
 \alpha +Q)+\alpha  (\alpha +3 Q)\right)\Big)~,
\\
D_2 &=\frac{1}{16} \bigg\{\Delta _{\gamma } \bigg(\big(1-2 (2 \alpha -3
 \beta_1 )^2\big) \Delta _{\gamma }-72 \beta_1 ^4-6 \beta_1 ^2 \big(10 \alpha
 ^2+6 Q^2+30 \alpha  Q-11\big)
\nonumber\\
&\qquad  +4 \beta_1  \big(2 \alpha ^3+3 \alpha  \big(5 Q^2-6\big)+21 \alpha ^2
 Q-6 Q\big)-8 \alpha  \big(Q \big(\alpha ^2+3 \alpha  Q-3\big)-2 \alpha
 \big)
\nonumber\\
&\qquad +12 \beta_1 ^3 (10 \alpha +9 Q)-1\bigg)-(Q-\beta_1 ) \bigg(3 \beta_1  \big(-24
 \beta_1 ^4+5 \beta_1 ^2 \big(7-6 Q^2\big)+\beta_1  Q \big(6 Q^2-35\big)
\nonumber\\
&\qquad +14 Q^2+48 \beta_1 ^3 Q-1\big)+6 \alpha ^2 \big(-16 \beta_1 ^3+8 \beta_1 +3
 Q^3-17 \beta_1  Q^2+5 \big(6 \beta_1 ^2-1\big) Q\big)
\nonumber\\
&\qquad -2 \alpha  \big(-72 \beta_1 ^4+70 \beta_1 ^2+18 \beta_1  Q^3+\big(21-90 \beta_1
 ^2\big) Q^2+4 \big(36 \beta_1 ^2-17\big) \beta_1  Q-1\big)
\nonumber\\
& \qquad +2 \alpha ^4 (Q-\beta_1 )+4 \alpha ^3 (3 (Q-2 \beta_1 )
 (Q-\beta_1 )-1)\bigg)\bigg\}~.
\label{eq:Q-dependent-D4}
\end{align}
Some more coefficients for higher orders are shown in appendix \ref{app:detail-D4}. 

Let us now consider the prefactor $ \langle \Delta_\gamma |
I^{(1)}(\beta_1,c_1;\beta_0)\rangle$ in \eqref{eq:result-D4}. Here
$|I^{(1)}(\beta_1,c_1;\beta_0)\rangle$ is a rank-one irregular state
satisfying \eqref{eq:irregular} for $n=1$. In the case of
$\beta_0=\beta_1$, the $c_1$-expansion of
$|I^{(1)} (\beta_1,c_1;\beta_0)\rangle$ was
studied in \cite{Gaiotto:2009ma}, whose generalization to $\beta_0\neq
\beta_1$ is of the form
\begin{align}
 |I^{(1)}(\beta_1,c_1;\beta_0)\rangle =
 c_1^{\Delta_{\beta_1}-\Delta_{\beta_0}}\left(|\Delta_{\beta_0}\rangle + \sum_{k=1}^\infty
 (c_1)^k |R_{k}\rangle\right)~,
\label{eq:beta1}
\end{align}
where the $|R_{k}\rangle$ is a level-$k$ Virasoro descendant of $|\Delta_{\beta_0}\rangle$.\footnote{Note here that the expression \eqref{eq:beta1}
fixes the ``boundary condition'' for
$|I^{(1)}(\beta_1,c_1;\beta_0)\rangle$ at $c_1=0$.}
While the Virasoro descendants $|R_k\rangle$ are fixed by
$\beta_0$ and $\beta_1$ so that $|I^{(1)}(\beta_1,c_1;\beta_0)\rangle$
satisfies \eqref{eq:irregular}, we do not need to evaluate them
explicitly. Indeed, since $|R_k\rangle$
are all Virasoro descendants,
we see that
\begin{align}
\langle \Delta_\gamma | I^{(1)}(\beta_1,c_1;\beta_0)\rangle = \left\{
\begin{array}{l}
c_1^{\Delta_{\beta_1}-\Delta_{\beta_0}}\quad (\Delta_{\beta_0} = \Delta_\gamma)\\
0\qquad\quad  (\Delta_{\beta_0} \neq \Delta_\gamma)\\
\end{array}
\right.~.
\label{eq:c1}
\end{align}
For this inner product to be non-vanishing, we set $\gamma=\beta_0$ in
the rest of this sub-section.

Substituting \eqref{eq:c1} in \eqref{eq:result-D4}, we obtain an explicit
expression for $\langle
\Delta_{\beta_0}|I^{(2)}(\alpha,\pmb{c};\pmb{\beta})\rangle$. This is
then identified via \eqref{eq:AGT-AD} with the partition function of the
$(A_1,D_4)$ theory, i.e.,
\begin{align}
\mathcal{Z}_{(A_1,D_4)}^{\epsilon_1,\epsilon_2} = c_1^{2(\alpha-\beta_1)(Q-\beta_1)+\Delta_{\beta_0}-\Delta_\alpha}c_2^{\frac{1}{2}(\beta_1-\alpha)(3Q-3\beta_1-\alpha)}e^{(\alpha-\beta_1)\frac{c_1^2}{c_2}}\sum_{k=0}^\infty \left(\frac{c_2}{ic_1^2}\right)^kD_k(\alpha,\beta_1,\beta_0)~,
\label{eq:final-D4}
\end{align}
up to a $c_i$-independent prefactor, where the  coefficients $D_i$ are
those in \eqref{eq:Q-dependent-D4-D0} -- \eqref{eq:Q-dependent-D4} with
$\gamma=\beta_0$. Recall here that $\sqrt{ic_1^2/c_2}$ is identified
with the relevant coupling of dimension $1/2$ of the $(A_1,D_4)$
theory. Therefore, \eqref{eq:final-D4} is an expansion of the $(A_1,D_4)$
partition function in the {\it inverse} powers of the relevant coupling, i.e.,
the strong coupling expansion of $\mathcal{Z}_{(A_1,D_4)}^{\epsilon_1,\epsilon_2}$.

Let us now compare \eqref{eq:final-D4} with 
the $(A_1,D_4)$ partition function obtained in
\cite{Bonelli:2016qwg}.\footnote{In this reference, the $(A_1,D_4)$
theory is called the $H_2$ Argyres-Douglas theory.} Since the expression
obtained in
\cite{Bonelli:2016qwg} is for the case $Q=0$, we will focus on the $Q\to
0$ limit of \eqref{eq:final-D4}
in the rest of this sub-section. Note that $Q=0$ corresponds to $\epsilon_1 +
\epsilon_2 = 0$ in four dimensions. We
first change the variables from
$(c_2/ic_1^2,\alpha,\beta_0,\beta_1)$ to
$(s,\nu,\theta_s,\theta_t)$ by
 \begin{align}
\frac{c_2}{ic_1^2}\equiv \frac{1}{s}~,\qquad \alpha \equiv
  i(\theta_s+\theta_t)~,\qquad \beta_0 \equiv i(\theta_t-\theta_s)~, \qquad \beta_1 \equiv \frac{i(\theta_s+\theta_t-3\nu)}{3}~.
\label{a1d4para}
\end{align}
Note that these relations could receive $\mathcal{O}(Q)$ corrections, 
as we will discuss in Sec.~\ref{sec:flavor}. We
omit such $\mathcal{O}(Q)$ corrections for the moment since we 
here focus on the limit $Q\to 0$. 
With the above change of variables, the partition function
for $Q=0$ is written as
\begin{align}
 \mathcal{Z}_{(A_1,D_4)}^{\epsilon_1,\epsilon_2}\big|_{\epsilon_1+\epsilon_2=0}
 &= c_1^{-\frac{(3\nu+2(\theta_s+\theta_t))^2}{9}
} 
c_2^{-\frac{3\nu^2}{2}+
 \frac{2(\theta_s+\theta_t)^2}{3}
}e^{\nu s +
 \frac{2}{3}(\theta_s+\theta_t)s}\sum_{k=0}^\infty \frac{D_k\big|_{Q=0}}{s^k}~.
\label{eq:Q=0-D4}
\end{align}
We see that
 this expression is identical to
$\mathcal{G}(\nu,s)$ in Eq.~(3.48) of \cite{Bonelli:2016qwg} obtained via the
connection to Painlev\'e IV, up to an
$s$-independent prefactor. In particular, the asymptotic behavior,
$e^{\nu s + \frac{2}{3}(\theta_s+\theta_t)s}$, in the
limit $s\to \infty$ is precisely identical. Moreover, 
our $D_1|_{Q=0}$ and $D_2|_{Q=0}$ are written as
\begin{align}
D_1\big|_{Q=0} &= 3\nu^3-2 \nu  \left(\theta_s^2-\theta_s
 \theta_t+\theta_t^2\right)-\frac{2}{9} (2 \theta_s-\theta_t)
 (\theta_s+\theta_t) (\theta_s-2 \theta_t)~,
\label{eq:D1-D4}\\
D_2\big|_{Q=0} &= \frac{9 \nu ^6}{2} +
\nu ^4 \Big(\frac{105}{16}-6 (\theta_s^2-\theta_s \theta_t+\theta_t^2)\Big)-\frac{2}{3} \nu ^3 (\theta_s-2 \theta_t) (2 \theta_s-\theta_t) (\theta_s+\theta_t)
\nonumber\\
&\qquad +\nu ^2 \Big(\frac{1}{2} (\theta_s^2-\theta_s \theta _t+\theta _t^2) (4 (\theta_s^2-\theta_s \theta_t+\theta _t^2)-11)+\frac{3}{16}\Big)
\nonumber\\
&\qquad +\frac{\nu}{9}  (2 \theta_s-\theta_t) (\theta_s+\theta_t) (\theta_s-2 \theta_t) (4 (\theta_s^2-\theta_s \theta_t+\theta _t^2)-7)
\nonumber\\
&\qquad +\frac{1}{324} \Big(32 \theta_s^6-96 \theta_s^5 \theta _t+\theta_s^4 (36-24 \theta_t^2)
+8 \theta_s^3 \theta_t (26 \theta_t^2-9)-3 \theta_s^2 (8 \theta _t^4-36 \theta_t^2+9)
\nonumber\\
&\qquad \qquad -3 \theta_s \theta_t (32 \theta _t^4+24
 \theta_t^2-9)+\theta_t^2 (32 \theta_t^4+36 \theta_t^2-27)\Big)~,
\label{eq:D2-D4}
\end{align}
which are precisely identical to
$D_1$ and $D_2$ in Eq.~(3.49) of \cite{Bonelli:2016qwg}. Therefore, the $Q\to 0$
limit of our partition function
$\mathcal{Z}_{(A_1,D_4)}^{\epsilon_1,\epsilon _2}$ is identical to
$\mathcal{G}(\nu,s)$ in \cite{Bonelli:2016qwg} up to a
prefactor. In particular, the strong coupling expansion 
in powers of $1/s$ corresponds to the $c_2$-expansion of
the irregular state $|I^{(2)}(\alpha,\pmb{c};\pmb{\beta})\rangle$ arising from
the ansatz \eqref{eq:expansion-D4}.

Note here that the above $Q\to 0$ limits of $D_1$ and $D_2$ were also
obtained in \cite{Nagoya:2015cja}, with a different parameterization of
the irregular conformal block.
In particular, the expansion parameter $1/s$ is identified in
\cite{Nagoya:2015cja} as the coordinate
of the vertex operator corresponding to the regular singularity.  
On the other hand, we here identify the expansion in powers of $1/s$ as the
asymptotic $c_2$-expansion arising from the ansatz \eqref{eq:expansion-D4} proposed in
\cite{Gaiotto:2012sf}.
These two parameter identifications are expected to be related by a
change of
coordinate on sphere.
We have also evaluated the $Q$-dependent
terms as shown in \eqref{eq:Q-dependent-D4}, which will be important in
our discussion in Sec.~\ref{sec:flavor}.





\subsection{$(A_1,A_3)$ theory from rank-3 irregular state}
\label{sec:A1A3}


Let us now turn to the $(A_1, A_3)$ theory. According to the generalized
AGT correspondence, the 
partition function of the $(A_1,A_3)$ theory is expected to be given by
\begin{align}
Z^{\epsilon_1,\epsilon_2}_{(A_1,A_3)} &= \langle 0|I^{(3)}(\alpha,\pmb{c};\pmb{\beta})\rangle~,
\label{eq:inner2}
\end{align}
where $|0\rangle$ is the vacuum state, and
$|I^{(3)}(\alpha,\pmb{c};\pmb{\beta})\rangle$ is an irregular primary
state of rank-three. Here
$\pmb{c}\equiv(c_1,c_2,c_3)$ and $\pmb{\beta}\equiv(\beta_0,\beta_1,\beta_2)$. This irregular state satisfies the following equations:
\begin{align}
L_6|I^{(3)}(\alpha,\pmb{c};\pmb{\beta})\rangle &= -c_3^2|I^{(3)}(\alpha,\pmb{c};\pmb{\beta})\rangle~,
\label{eq:L6-A3}
\\
L_5 |I^{(3)}(\alpha,\pmb{c};\pmb{\beta})\rangle &= -2c_2c_3|I^{(3)}(\alpha,\pmb{c};\pmb{\beta})\rangle~,
\\
L_4 |I^{(3)}(\alpha,\pmb{c};\pmb{\beta})\rangle &= -\left(c_2^2 + 2c_3c_1\right)|I^{(3)}(\alpha,\pmb{c};\pmb{\beta})\rangle~,
\\
L_3 |I^{(3)}(\alpha,\pmb{c};\pmb{\beta})\rangle &= -2\Big(c_1c_2 + c_3(\alpha-2Q)\Big)|I^{(3)}(\alpha,\pmb{c};\pmb{\beta})\rangle~,
\\
L_2 |I^{(3)}(\alpha,\pmb{c};\pmb{\beta})\rangle &= \left(c_3\frac{\partial}{\partial c_1} -c_2(2\alpha-3Q)-c_1^2\right)|I^{(3)}(\alpha,\pmb{c};\pmb{\beta})\rangle~,
\\
L_1 |I^{(3)}(\alpha,\pmb{c};\pmb{\beta})\rangle &= \left(2c_3\frac{\partial}{\partial c_2}+c_2\frac{\partial}{\partial c_1}-2c_1(\alpha-Q)\right)|I^{(3)}(\alpha,\pmb{c};\pmb{\beta})\rangle~,
\\
L_0|I^{(3)}(\alpha,\pmb{c};\pmb{\beta})\rangle &= \left(\Delta_\alpha + c_1\frac{\partial}{\partial c_1} + 2c_2\frac{\partial}{\partial c_2}+3c_3\frac{\partial}{\partial c_3}\right)|I^{(3)}(\alpha,\pmb{c};\pmb{\beta})\rangle~.
\label{eq:L0-A3}
\end{align}

The relation between the 4d and 2d parameters is identified by looking
at the Seiberg-Witten curve. Indeed, $T(z) =
\sum_{n=-\infty}^{\infty}L_nz^{-n-2}$ implies that 
\begin{align}
\frac{\langle 0|\,T(z)\,|I^{(3)}(\alpha,\pmb{c};\pmb{\beta})\rangle}{\langle 0|I^{(3)}(\alpha,\pmb{c};\pmb{\beta})\rangle} &= \frac{u}{z^4} -\frac{2\Big(c_1c_2 + c_3(\alpha-2Q)\Big)}{z^5} -\frac{c_2^2+2c_3c_1}{z^6}-\frac{2c_2c_3}{z^7}-\frac{c_3^2}{z^8}~,
\end{align}
where $u \equiv c_3 \frac{\partial}{\partial c_1}\log\, \langle 0|I^{(3)}(\alpha,c_1,c_2,c_3)\rangle -c_2(2\alpha-3Q)-c_1^2$.
By changing the variable as $z\to -i c_3^{1/3}z/\big(1+ic_2z/2c_3^{2/3}\big)$, the stress
tensor changes as $T \to
-c_3^{-2/3}\big(1+ic_2z/2c_3^{2/3}\big)T$.\footnote{Note that this
transformation is generated by $L_0$ and $L_{-1}$, and therefore the
stress tensor transforms as a tensor under it.} Then the above
equation is mapped to
\begin{align}
\frac{\langle
 0|\,T(z)\,|I^{(3)}(\alpha,\pmb{c};\pmb{\beta})\rangle}{\langle
 0|I^{(3)}(\alpha,\pmb{c};\pmb{\beta})\rangle} &= \frac{\tilde{u}}{z^4}
 +\frac{2i(2Q-\alpha)}{z^5}
 +\frac{\frac{(c_2)^2}{(c_3)^{4/3}}-\frac{4c_1}{(c_3)^{1/3}}}{2z^6}+\frac{1}{z^8}~,
\end{align}
where $\tilde{u}\equiv -u/(c_3)^{2/3}+\big(-8 c_1 (c_2)^2 c_3+(c_2)^4+16 c_2 (c_3)^2 (\alpha -2 Q)\big)/16 (c_3)^{8/3}$.
Using \eqref{eq:curve-from-2d}, we can relate this to the Seiberg-Witten curve
of the $(A_1,A_3)$ theory shown in \eqref{eq:curve-A3}. To be concrete,
$\tilde u$ is identified with the vev of a Coulomb branch operator of
dimension $4/3$, $2i(2Q-\alpha)$ is the mass parameter, and
$(c_2)^2/(c_3)^{4/3} - 4c_1/(c_3)^{1/3}$ is identified with the relevant coupling of dimension $2/3$. Note that
while the irregular state $|I^{(3)}(\alpha,\pmb{c};\pmb{\beta})\rangle$ depends
on $c_1,c_2$ and $c_3$, the Seiberg-Witten curve depends only on the
above particular combination of them. This means that there is a 
redundancy in describing the partition function of the $(A_1,A_3)$ theory in
terms of the 2d irregular conformal block.
We use this redundancy to set
$c_1=0$ in the rest of this section, which turns out to simplify the
computation. Under the condition $c_1=0$, the 4d relevant
coupling of dimension $2/3$ is identified with 
\begin{align}
\frac{(c_2)^2}{(c_3)^{4/3}} = \left(\frac{(c_2)^3}{(c_3)^2}\right)^{\frac{2}{3}}~.
\end{align}

\subsubsection{Irregular conformal block}

Let us now evaluate the rank-three irregular state
$|I^{(3)}(\alpha,\pmb{c};\pmb{\beta})\rangle$ explicitly.
To that end,  we first use the ansatz \eqref{eq:general-ansatz} for the
rank-three state:
\begin{align}
|I^{(3)}(\alpha,\pmb{c};\pmb{\beta})\rangle &=
 c_2^{\rho_2}c_3^{\rho_3}e^{(\alpha-\beta_2)S_3(\pmb{c})}\left(|I^{(2)}(\beta_2,\tilde{\pmb{c}};\tilde{\pmb{\beta}})\rangle
 + \sum_{k=1}^\infty (c_3)^k |I^{(2)}_{k}\rangle\right)~,
\label{eq:expansion-A3}
\end{align}
where $S_3(\pmb{c})\equiv \frac{2c_1c_2}{c_3} - \frac{c_2^3}{3c_3^2} -
 \frac{c_1^2}{c_2}$, $\tilde{\pmb{c}} \equiv (c_1,c_2)$,
 $\tilde{\pmb{\beta}}\equiv (\beta_0,\beta_1)$, and $|I^{(2)}_{k}\rangle$ are
 generalized descendants of the rank-two irregular state
 $|I^{(2)}(\beta_2,\tilde{\pmb{c}};\tilde{\pmb{\beta}})\rangle$.\footnote{As mentioned
 already, our convention for $|I^{(2)}_k\rangle$ is slightly different
 from that of \cite{Gaiotto:2012sf}. Our $|I^{(2)}_k\rangle$ is
 identical to $|I^{(2)}_{3k}\rangle$ of \cite{Gaiotto:2012sf}.} The
 generalized descendants are now of the form
\begin{align}
 (L_{-\ell_1})^{p_1}(L_{-\ell_2})^{p_2}\cdots(L_{-\ell_n})^{p_n}\frac{\partial
^{m_1}}{\partial c_1{}^{m_1}}\frac{\partial^{m_2}}{\partial c_2{}^{m_2}}|I^{(2)}(\beta_2,\tilde{\pmb{c}};\tilde{\pmb{\beta}})\rangle~,
\label{eq:I2general}
\end{align}
for integers $n,m_i\geq 0$,\, $p_i>0$ and $\ell_1>\ell_2>\cdots>\ell_n>0$.
 The exponents, $\rho_2$ and $\rho_3$, are uniquely determined as
\begin{align}
\rho_2 = \frac{1}{2}(\beta_2-\alpha)(5Q-5\alpha+\beta_2)~,\qquad
\rho_3 = \frac{2}{3}(\alpha-\beta_2)(2Q-2\alpha+\beta_2)~,
\label{eq:rho2and3}
\end{align}
so that the above ansatz solves \eqref{eq:L6-A3} --
\eqref{eq:L0-A3}. Note that the ansatz \eqref{eq:expansion-A3} again
fixes the ``asymptotic behavior'' of
$|I^{(3)}(\alpha,\pmb{c};\pmb{\beta})\rangle$ in the limit $c_3\to0$. The generalized
descendants $|I^{(2)}_{k}\rangle$ are conjectured in
\cite{Gaiotto:2012sf} to be uniquely determined. We
explicitly computed them for $k=1,\cdots,6$, whose first several
expressions are shown in appendix \ref{app:A3-detail}. 

Using the expressions for the generalized descendants, we see that
the inner product of \eqref{eq:expansion-A3} and $|0\rangle$ is
evaluated as
\begin{align}
\langle 0|I^{(3)}(\alpha,\pmb{c};\pmb{\beta})\rangle &= \langle 0 |
 I^{(2)}(\beta_2,\tilde{\pmb{c}};\tilde{\pmb{\beta}})\rangle
 c_2^{\rho_2}c_3^{\rho_3}e^{(\alpha-\beta_2)S_3(\pmb{c})}\left(\sum_{k=0}^\infty
 \left(\frac{3(c_3)^2}{(c_2)^3}\right)^k \!\! D_k(\alpha,\beta_2) + \mathcal{O}(c_1)\right)~,
\label{eq:pre-final-A3}
\end{align}
where
\begin{align}
D_0 &= 1~,
\label{eq:D0-A3-Q}
\\[1mm]
D_1 &= -\frac{1}{36} \left(\alpha -2 \beta _2+Q\right) \left(4 \alpha
 ^2+34 \beta _2^2-34 \beta _2 (\alpha +Q)+35 \alpha  Q-1\right)~,
\label{eq:D1-A3-Q}
\\[2mm]
D_2 &= \frac{1}{2592}(16 \alpha ^6+\alpha ^4 \big(1801 Q^2-200\big)+2 \alpha
 ^3 Q \big(1365 Q^2-1489\big)+50 \alpha  Q \big(19-77 Q^2\big)
\nonumber\\
&\qquad -35
 Q^2+\alpha ^2 \big(1225 Q^4-6244 Q^2+169\big)-4 \beta _2
 \big(\alpha -\beta _2+Q\big) \big(84 \alpha ^4-763 \alpha ^2
\nonumber\\
&\qquad +595
 \alpha  Q^3-\beta _2 \big(\alpha -\beta _2+Q\big) \big(561 \alpha
 ^2+289 Q^2-1156 \beta _2 \big(\alpha -\beta _2+Q\big)+2958 \alpha
 Q
\nonumber\\
&\qquad -2318\big)+\big(2483 \alpha ^2-971\big) Q^2+\big(1011 \alpha
 ^2-3902\big) \alpha  Q+271\big)+312 \alpha ^5 Q)~,
\\[2mm]
D_3 &= -\frac{1}{279936}\big(\alpha -2 \beta _2+Q\big) \bigg(64
 \alpha ^8-2352 \alpha ^6+26796 \alpha ^4-48313 \alpha ^2+42875 \alpha
 ^3 Q^5
\nonumber\\
&\qquad +1225 \big(82 \alpha ^2-327\big) \alpha ^2 Q^4+35 \big(2113
 \alpha ^4-19818 \alpha ^2+33843\big) \alpha  Q^3+\big(18124 \alpha ^6
\nonumber\\
&\qquad -382215 \alpha ^4+1420026 \alpha ^2-5005\big) Q^2-2 \beta _2
 \big(\alpha -\beta _2+Q\big) \big(944 \alpha ^6-28008 \alpha ^4
\nonumber\\
&\qquad +243975 \alpha ^2+62475 \alpha ^2 Q^4+70 \big(3214 \alpha
 ^2-5667\big) \alpha  Q^3+3 \big(40417 \alpha ^4-382090 \alpha ^2
\nonumber\\
&\qquad +196601\big) Q^2-2 \beta _2 \big(\alpha -\beta _2+Q\big)
 \big(5100 \alpha ^4-34 \beta _2 \big(\alpha -\beta _2+Q\big) \big(697
 \alpha ^2+289 Q^2
\nonumber\\
&\qquad -1156 \beta _2 \big(\alpha -\beta _2+Q\big)+4148 \alpha
 Q-6852\big)+9 \alpha ^2 \big(21012 Q^2-11551\big)
\nonumber\\
&\qquad +105 \alpha  Q \big(289 Q^2-6024\big)-98175 Q^2+65841 \alpha ^3
 Q+419196\big)+6 \big(3212 \alpha ^4
\nonumber\\
&\qquad -66615 \alpha ^2+256546\big) \alpha  Q-224894\big)+\big(1808
 \alpha ^6-55800 \alpha ^4+517557 \alpha ^2
\nonumber\\
&\qquad -481430\big) \alpha  Q+7560\bigg)~.
\label{eq:D3-A3-Q}
\end{align}
The prefactor $\langle 0 |
I^{(2)}(\beta_2,\tilde{\pmb{c}};\tilde{\pmb{\beta}})\rangle$ in
\eqref{eq:pre-final-A3} is
evaluated as follows. Since $\langle 0|L_1 = \langle 0|L_0 = 0$,
\eqref{eq:L1-D4} and \eqref{eq:L0-D4} imply
\begin{align}
\left(c_2\frac{\partial}{\partial c_1} -2c_1(\beta_2-Q)\right) \langle 0
 | I^{(2)}(\beta_2,\tilde{\pmb{c}};\tilde{\pmb{\beta}})\rangle =
 \left(\Delta_{\beta_2} + c_1\frac{\partial}{\partial c_1} +
 2c_2\frac{\partial}{\partial c_2}\right)\langle 0
 | I^{(2)}(\beta_2,\tilde{\pmb{c}};\tilde{\pmb{\beta}})\rangle =0~.
\end{align}
From this set of differential equations, we see that $\langle 0|I^{(2)}(\beta_2,\tilde{\pmb{c}},\tilde{\pmb{\beta}})\rangle $ is given by
\begin{align}
\langle 0|I^{(2)}(\beta_2,\tilde{\pmb{c}},\tilde{\pmb{\beta}})\rangle 
= e^{-(Q-\beta_2)\frac{c_1^2}{c_2}}c_2^{-\frac{1}{2}\Delta_{\beta_2}}~,
\end{align}
up to a prefactor independent of $c_1$ and $c_2$. Note here that $\tilde{\pmb{\beta}}$-dependence only
appears in the $c_i$-independent prefactor.\footnote{For this
prefactor to be non-vanishing, we need to set $\Delta_{\beta_0} = 0$. Indeed,
$|I^{(2)}(\alpha,\pmb{c};\pmb{\beta})\rangle$ is a linear combination of
$|I^{(1)}(\beta_1,c_1;\beta_0)\rangle$ and its generalized
descendants, and $|I^{(1)}(\beta_1,c_1;\beta_0)\rangle$ is a linear
combination of $|\Delta_{\beta_0}\rangle$ and its Virasoro
descendants. Therefore, $|I^{(2)}(\alpha,\pmb{c};\pmb{\beta})\rangle$ is
in the Verma module of Virasoro algebra whose highest weight state is
$|\Delta_{\beta_0}\rangle$. This means that $\langle 0|
I^{(2)}(\alpha,\pmb{c};\pmb{\beta})\rangle=0$ unless $\Delta_{\beta_0}= 0$.}

According to \eqref{eq:AGT-AA}, the partition
function of the $(A_1,A_3)$ theory is identified with $\langle
0|I^{(3)}(\alpha,\pmb{c};\pmb{\beta})\rangle$.
In this identification, we can set $c_1=0$,
as discussed at the beginning of this sub-section.
Therefore, we have
\begin{align}
\mathcal{Z}_{(A_1,A_3)}^{\epsilon_1,\epsilon_2}&= \langle 0
 |I^{(3)}(\alpha,\pmb{c};\pmb{\beta}) \rangle\big|_{c_1=0} 
=
 c_2^{-\frac{1}{2}\Delta_{\beta_2}+\rho_2}c_3^{\rho_3}e^{-(\alpha-\beta_2)\frac{c_2^3}{3c_3^2}}\sum_{k=0}^\infty
 \left(\frac{3c_3^2}{c_2^3}\right)^k D_k(\alpha,\beta_2)~,
\label{eq:final-A3}
\end{align}
up to a $c_i$-independent prefactor, where the first several
coefficients $D_k$ are shown in \eqref{eq:D0-A3-Q} -- \eqref{eq:D3-A3-Q}.
Recall here that $((c_2)^3/(c_3)^2\big)^{\frac{2}{3}}$ is identified with the
relevant coupling of dimension $2/3$ in the $(A_1,A_3)$
theory. Therefore \eqref{eq:final-A3} is an expansion of the $(A_1,A_3)$
partition function in the inverse powers of the coupling, i.e., 
the strong coupling expansion.

Let us now compare the $Q\to 0$ limit of \eqref{eq:final-A3} with the
strong coupling expansion of the $(A_1,A_3)$ partition function
evaluated in \cite{Bonelli:2016qwg}.\footnote{In this reference, the
$(A_1,A_3)$ theory is called the $H_1$ Argyres-Douglas theory.} To
that end, we first change the variables as
\begin{align}
\frac{3(c_3)^2}{(c_2)^3} \equiv \frac{1}{s}~,\qquad \alpha \equiv i\theta~,\qquad \beta_2 \equiv i\left(\nu+\frac{\theta}{2}\right)~.
\label{a1a3para}
\end{align}
Here, these relations could receive $\mathcal{O}(Q)$
corrections, as we will discuss in Sec.~\ref{sec:flavor}. We omit
such $\mathcal{O}(Q)$ corrections for a while since we focus on the $Q\to 0$ limit here.
With the above change of variables, the partition function for $Q=0$
(or equivalently $\epsilon_1+\epsilon_2=0$) is written as
\begin{align}
 \mathcal{Z}_{(A_1,A_3)}^{\epsilon_1,\epsilon_2}\big|_{\epsilon_1+\epsilon_2=0} &=  c_2^{-\frac{5
 }{4}\theta ^2 + 2 \theta  \nu -\nu ^2
}c_3^{\frac{1}{6} (\theta -2 \nu )
 (3 \theta -2 \nu ) 
}e^{i\left(\nu-\frac{\theta}{2}\right)s}\sum_{k=0}^\infty
 \frac{D_k\big|_{Q=0}}{s^k}~,
\label{eq:Q=0-A3}
\end{align}
where 
\begin{align}
D_1\big|_{Q=0} &= -\frac{17}{9}i\nu^3+ \frac{9\theta^2-2}{36}i\nu~,
\label{eq:D1-A3}
\\[2mm]
D_2\big|_{Q=0} &= -\frac{289}{162}\nu^6 +
 \frac{153\theta^2-1159}{324}\nu^4-\Big(\frac{\theta^4}{32}-\frac{11\theta^2}{18}+\frac{271}{648}\Big)\nu^2-\frac{\theta^2(11\theta^2-68)}{1728}~,
\label{eq:D2-A3}
\\[2mm]
D_3\big|_{Q=0} &= \frac{4913}{4374}i\nu^9 -\frac{17  \left(153 \theta ^2-2284\right)}{5832}i\nu^7+\frac{
 1377 \theta ^4-47178 \theta ^2+279464}{23328}i\nu^5
\nonumber\\[2mm]
&\qquad +\frac{-729 \theta ^6+45648 \theta ^4-700884 \theta
 ^2+899576}{279936}i\nu^3
 \nonumber\\[2mm]
 &\qquad \qquad
 +\frac{-99 \theta ^6+4270
 \theta ^4-28504 \theta ^2+3360}{62208}i\nu~.
\label{eq:D3-A3}
\end{align}
We now see that
the partition function \eqref{eq:Q=0-A3} for $Q=0$ coincides, up to a prefactor,
with $\mathcal{G}(\nu,s)$ in Eq.~(3.32) of \cite{Bonelli:2016qwg}
obtained via the connection to Painlev\'e II.
In particular, 
our $D_1|_{Q=0}$ and $D_2|_{Q=0}$ are
precisely identical to 
$D_1$ and $D_2$ in Eq.~(3.33) of
\cite{Bonelli:2016qwg}.\footnote{While the $D_3$ is not
explicitly written in \cite{Bonelli:2016qwg}, the authors of
\cite{Grassi:2018spf} evaluated it using a nice matrix model
description as shown in Eq.~(3.23) of \cite{Grassi:2018spf}. We then see that
the $Q\to 0$ limit of our $D_3$ is in perfect agreement with the
expression.} 
This
particularly means that the strong coupling expansion in powers of $1/s$
corresponds
to the asymptotic $c_3$-expansion of the rank-$3$ state
$|I^{(3)}(\alpha,\pmb{c};\pmb{\beta})\rangle$. 

Note here that \eqref{eq:D1-A3} and \eqref{eq:D2-A3} were also
reproduced from the irregular conformal block with
one irregular singularity of rank $\frac{3}{2}$ and one regular
singularity \cite{Nagoya:2018}.
On the other hand, we obtained the above results from the irregular conformal block with an
irregular singularity of rank $3$ without any regular singularity.
Therefore, the author of
\cite{Nagoya:2018} and we computed different
irregular conformal blocks of Liouville theory. Nevertheless, the two
irregular conformal blocks turn out to be related to the same
tau-function of Painlev\'e II. This is natural from the viewpoint of 4d/2d
correspondence. Indeed, the Hitchin systems associated with the two different irregular conformal blocks are
isomorhic \cite{Xie:2012hs}, and therefore these two
irregular conformal blocks are expected
to be related to 
the same 4d $\mathcal{N}=2$ SCFT, i.e., the $(A_1,A_3)$ theory.
Our discussion above explicitly confirms this expectation.


\section{
Enhanced flavor symmetries}
\label{sec:flavor}

Having evaluated the irregular conformal blocks for the $(A_1,D_4)$ and
$(A_1,A_3)$ theories, we here discuss their flavor symmetries. In particular, we
will show that the strong coupling expansions of
$\mathcal{Z}^{\epsilon_1,\epsilon_2}_{(A_1,D_4)}$ and $\mathcal{Z}^{\epsilon_1,\epsilon_2}_{(A_1,A_3)}$ are invariant under the action of
the Weyl group of the enhanced flavor symmetry
of the AD theories.

In general,
turning on generic values of mass parameters breaks the flavor
symmetry group to its maximal torus.\footnote{The reason for this is that, in any
4d $\mathcal{N}=2$ SCFT, every $\mathcal{N}=2$ preserving mass
deformation operator is in the same superconformal multiplet as a flavor
current, and therefore in the adjoint representation of the flavor
symmetry.} Therefore the complete flavor
symmetry is not visible in the partition function with mass parameters
turned on. However, the Weyl group of the flavor symmetry is still a
symmetry of the mass-deformed theory, since it is a symmetry of the
maximal torus. In particular, the action of the Weyl group permutes the mass parameters in such a
way that the partition function is invariant. Below, we identify such an
action of the Weyl
group for the
%
$(A_1,D_4)$ and $(A_1,A_3)$ theories. 

Furthermore, by demanding this Weyl group symmetry extend to the case of non-vanishing Liouville charge $Q$, we
propose an identification of parameters between the Loiuville
side and the Argyres-Douglas side for general values of the
$\Omega$-background parameters. Under this parameter identification,
we find that the $(A_1,D_4)$ and $(A_1,A_3)$ partition functions are
invariant under $Q\to -Q$, or equivalently
$(\epsilon_1,\epsilon_2)\to(-\epsilon_1,-\epsilon_2)$. We will also give
a natural explanation for our parameter identification from the AGT
viewpoint, which leads us to a conjecture on the parameter identification for a general rank-$n$ irregular state.



\subsection{Weyl group of the flavor symmetry 
}

We first study how the Weyl group of the enhanced flavor symmetry is visible in the
irregular conformal blocks corresponding to the $(A_1,D_4)$ and
$(A_1,A_3)$ theories.


\subsubsection{$(A_1,D_4)$ theory}

We start with the $(A_1,D_4)$
 partition function \eqref{eq:final-D4} evaluated as the irregular
 conformal block with one irregular
 singularity of rank $2$ and one regular singularity. The flavor symmetry of the $(A_1,D_4)$
 theory is enhanced from the manifest $\SU(2)\times \U(1)$ to $\SU(3)$,
 as reviewed in Sec.~\ref{sec:review}. Therefore its Weyl group $S_3$ is expected to act on the
 set of mass parameters so that the 
the $(A_1,D_4)$ partition function is invariant. 

Such an action of $S_3$ can easily be found in the limit of
$Q\to0$. To see this, let us look at the $Q\to 0$ limit of $\mathcal{Z}^{\epsilon_1,\epsilon_2}_{(A_1,D_4)}$ shown in
\eqref{eq:Q=0-D4}. Here, explicit expressions for $D_k|_{Q=0}$ are shown in \eqref{eq:D1-D4}
-- \eqref{eq:D2-D4}, where $\theta_s,\theta_t$ and $\nu$ are related by
\eqref{a1d4para} to
the parameters $\alpha,\beta_0$ and $\beta_1$ of the corresponding irregular conformal
block. As shown in \cite{Bonelli:2016qwg}, $\theta_s$ and
$\theta_t$ are identified with two mass parameters of the
$(A_1,D_4)$ theory. 
We now see that the expansion coefficients $D_1|_{Q=0}$ and $D_2|_{Q=0}$ are
invariant under both of the following two transformations
\begin{align}
\sigma_1 &: \;\; \theta_s \longleftrightarrow \theta_t~,
\label{a1d4Weyl1}
\\[1mm]
\sigma_2 &:\;\; \theta_s\rightarrow\theta_s-\theta_t~,\quad \theta_t\rightarrow-\theta_t~.
\label{a1d4Weyl2}
\end{align}
Note that the above $\sigma_1$ and $\sigma_2$ satisfy 
\begin{align}
 \sigma_1^2 = \sigma_2^2 = (\sigma_1\sigma_2)^3 = {\bf 1}~,
\end{align}
and therefore gives a representation of $S_3$. 
This $S_3$ is naturally identified with
the Weyl group of the flavor $\SU(3)$ symmetry of $(A_1,D_4)$.

Let us now consider the case of $Q\neq 0$. Recall that the Liouville
charge $Q$ corresponds to the
$\Omega$-background parameters by \eqref{eq:Omega}.  Since the
$\Omega$-deformation does not break the flavor symmetry, the
above $S_3$-invariance 
of the partition function is expected to extend to the case of $Q\neq
0$. 
Indeed, we find that 
$D_k$ in
\eqref{eq:final-D4} are invariant under the $S_3$ generated by
\eqref{a1d4Weyl1} and \eqref{a1d4Weyl2} if 
the parameter identification \eqref{a1d4para} receives an
$\mathcal{O}(Q)$-correction as 
\begin{align}
\alpha \equiv i(\theta_s+\theta_t)+\frac{3}{2}Q~,
\qquad 
 \beta_0 \equiv i(\theta_s-\theta_t)+\frac{1}{2}Q~,
\qquad 
\beta_1 = \frac{i(\theta_s+\theta_t-3\nu)}{3} + \mathcal{O}(Q)~,
\label{a1d4paraQ}
\end{align}
where the $\mathcal{O}(Q)$ correction to $\beta_1$ is arbitrary as far
as it is 
independent of $\theta_s$ and $\theta_t$.\footnote{The $\mathcal{O}(Q)$-term is
expected to be independent of $\theta_s$ and $\theta_t$, since the
$\Omega$-deformation is not coupled with a  mass deformation.} 
Note here that
the $S_3$ invariance requires non-vanishing $\mathcal{O}(Q)$ corrections to the
parameter identification \eqref{a1d4para}. With the modified identification
\eqref{a1d4paraQ}, we have checked that
$D_1,\cdots,D_4$ in \eqref{eq:final-D4} are invariant under the $S_3$
generated by \eqref{a1d4Weyl1} and \eqref{a1d4Weyl2}. See appendix
\ref{app:D3 with Q} for more detail.

The $\mathcal{O}(Q)$ correction to $\beta_1$ is not fixed by 
the $S_3$ invariance. The reason for this
is that $\beta_1$ is related to the parameter $\nu$ which is identified with the
(dual) Coulomb branch parameter $a_D$ by $\nu = ia_D$ (See the last
line of Sec.~4 of
\cite{Bonelli:2016qwg}). Since $a_D$ is neutral under the flavor
$\SU(3)$ symmetry, so is $\nu$. Then the
flavor symmetry is not enough to fix the $\mathcal{O}(Q)$ correction to the relation between
$\beta_1$ and $\nu$. Nevertheless, we will
conjecture in the next sub-section that the correct identification is  $\beta_1 =
i(\theta_s+\theta_t-3\nu)/3 + Q$.



\subsubsection{$(A_1,A_3)$ theory}

We now turn to the $(A_1,A_3)$ partition function evaluated as the irregular conformal block with one
irregular singularity of rank $3$. In the limit $Q\to 0$, the strong
coupling expansion of the partition
function is given by \eqref{eq:Q=0-A3}.
We see that $D_1|_{Q=0},\cdots,D_3|_{Q=0}$ in the expansion are invariant under the $S_2$-action
\begin{align}
\sigma:\theta\rightarrow-\theta~.
\label{a1a3Weyl}
\end{align}
Since $\theta$ is identified with the mass parameter associated with the
flavor $\SU(2)$ symmetry of the $(A_1,A_3)$ theory, it is natural to
identify this $S_2$ as the Weyl group of the flavor symmetry.

This $S_2$-invariance of the partition function can be extended
to the case of $Q\neq 0$, if the parameter identification
\eqref{a1a3para} receives an $\mathcal{O}(Q)$-correction as
\begin{align}
\alpha\equiv i\theta +2Q~,\qquad \beta_2 = i\left(\nu +
 \frac{\theta}{2}\right) + \mathcal{O}(Q)~,
\label{a1a3paraQ}
\end{align}
where the $\mathcal{O}(Q)$ correction to $\beta_2$ is arbitrary as far
as it is independent of $\theta$. We have checked that $D_1, D_2$ and
$D_3$ in \eqref{eq:D1-A3-Q} -- \eqref{eq:D3-A3-Q} are invariant under
the $S_2$-action
\eqref{a1a3Weyl}, under the modified parameter identification
\eqref{a1a3paraQ}. Note that 
the $S_2$-invariance of the partition function
requires a non-vanishing $\mathcal{O}(Q)$ correction to the parameter
identification. The $\mathcal{O}(Q)$ correction to the relation between
$\beta_2$ and $\nu$ is not fixed by the flavor symmetry, since $\nu$ is related to the (dual) Coulomb
branch parameter and therefore neutral under the flavor
symmetry. Nevertheless, the correction identification will be conjectured in
the next sub-section to be $\beta_2 = i\left(\nu + \theta/2\right) + 3Q/2$.





\subsection{Conjectural dictionary between 4d and 2d parameters
}

As seen in the previous sub-section,  
the strong coupling expansions of the $(A_1,D_4)$ and $(A_1,A_3)$
partition functions are invariant under the action of the Weyl group of the flavor
symmetries. This Weyl group symmetry arises if 
the parameter identifications, \eqref{a1d4para} and \eqref{a1a3para},
are modified by 
$\mathcal{O}(Q)$ corrections as in \eqref{a1d4paraQ} and
\eqref{a1a3paraQ}, respectively. This means that there is generally an
$\mathcal{O}(Q)$-correction to the relation between the mass parameters of
4d $\mathcal{N}=2$ theories and parameters in 2d Liouville theory.

To read off a general rule for this $\mathcal{O}(Q)$ correction, note first that $\alpha$ and $\beta_0$ in \eqref{a1d4paraQ} are
regarded as the Liouville momenta of a rank-$2$ 
state and
a 
rank-$0$ state, respectively. Indeed, 
$\alpha$ is the Liouville momentum of
 $|I^{(2)}(\alpha,\pmb{c};\pmb{\beta})\rangle$,\footnote{Recall here that we call $\alpha$ in
\eqref{eq:irregular} the ``Liouville momentum'' of the irregular state
of rank $n$.}
and $\beta_0$ is 
the Liouville momentum of 
the regular primary state $|\Delta_{\beta_0}\rangle$ 
that arises 
in the expansion of $|I^{(2)}(\alpha,\pmb{c};\pmb{\beta})\rangle$
 in terms of
 generalized descendants.\footnote{Recall
 that a rank-zero state is identified as a
 regular primary state.} 
Similarly, 
$\alpha$  in
 \eqref{a1a3paraQ} is the Liouville momentum of the rank-$3$
state $|I^{(3)}(\alpha,\pmb{c};\pmb{\beta})\rangle$.
More generally, as reviewed in Sec.~\ref{sec:review}, the
 parameters $\alpha$ and $\beta_k$ of
 $|I^{(n)}(\alpha,\pmb{c};\pmb{\beta})\rangle$ are respectively regarded as the
 Liouville momenta of rank-$n$ and rank-$k$ states,
 in the expansion in terms of generalized descendants. This follows from the general ansatz
 \eqref{eq:general-ansatz} for
 $|I^{(n)}(\alpha,\pmb{c};\pmb{\beta})\rangle$ proposed in
 \cite{Gaiotto:2012sf}.

With this correspondence in mind, we now see that the first two equations of \eqref{a1d4paraQ} and the
first equation
of \eqref{a1a3paraQ} satisfy the following rule: {\it when the Liouville
momentum $\alpha$ of a rank-$n$ state is associated with a
four-dimensional mass parameter (or the vev of a Coulomb branch parameter) $m$, the precise relation between them is given by}
\begin{align}
 \alpha = m + \frac{n+1}{2}Q~.
\label{eq:dictionary}
\end{align}
Here $m$ is a  mass parameter or the vev of a (dual) Coulomb
branch operator, depending on whether the corresponding symmetry is gauged.

Note that the above rule \eqref{eq:dictionary} is a natural
generalization of the original AGT correspondence involving only regular
primary states \cite{Alday:2009aq}. In \cite{Alday:2009aq}, it was shown that
the Liouville momentum $\gamma$ of a regular primary state is related to
a mass parameter $m$ of the four-dimensional theory by $\gamma = m +
Q/2$. Since the regular state is regarded as the rank-zero state, this
is a special case of \eqref{eq:dictionary}.
The
factor $\frac{n+1}{2}$ in \eqref{eq:dictionary} is understood as follows. Recall that the
irregular state of rank $n$ is constructed by taking a colliding limit
of $(n+1)$ regular primary states. As reviewed in Sec.~\ref{sec:review},
the Liouville momentum $\alpha$ of the resulting rank-$n$ 
state $|I^{(n)}(\alpha,\pmb{c};\pmb{\beta})\rangle$ is the
sum of the Liouville momenta of the original $(n+1)$ regular primary
states, say $\gamma_0,\cdots,\gamma_{n}$. On the other hand, the original
AGT correspondence implies that the $i$-th momentum $\gamma_i$ is related to a
four-dimensional mass parameter $m_i$ by $\gamma_i = m_i +
Q/2$. Combining these two, we see that 
\begin{align}
 \alpha = \sum_{i=0}^{n}\gamma_i = \left(\sum_{i=0}^{n}m_i\right) +
 \frac{n+1}{2}Q~.
\label{eq:colliding}
\end{align}
By identifying $\sum_{i=0}^{n}m_i$ as the 4d mass parameter $m$ corresponding
to the Liouville momentum $\alpha$, we see that \eqref{eq:colliding} is
precisely identical to the rule \eqref{eq:dictionary}.

Given the above interpretation of \eqref{eq:dictionary}, we now
conjecture that the rule \eqref{eq:dictionary} generally applies to the
Liouville momentum $\alpha$ of any
Liouville irregular state of integer rank $n$ and the corresponding 4d mass
parameter $m$.\footnote{There are also
irregular states of half-integer rank. They are not constructed in a
colliding limit of regular primary states, and moreover they have no
parameter corresponding to a 4d mass parameter.} This particularly implies that the
$\mathcal{O}(Q)$ correction to $\beta_1$ in \eqref{a1d4paraQ} is
fixed by
\begin{align}
 \beta_1 = \frac{i(\theta_s+\theta_t-3\nu)}{3} + Q~,
\label{a1d4-extra}
\end{align}
 and
that to $\beta_2$ in \eqref{a1a3paraQ} is fixed by
\begin{align}
  \beta_2 = i\left(\nu+ \frac{\theta}{2}\right) + \frac{3}{2}Q~.
\label{a1a3-extra}
\end{align}
Below, we rewrite the strong coupling expansions of the $(A_1,D_4)$ and
$(A_1,A_3)$ partition functions using \eqref{a1d4paraQ} and
\eqref{a1a3paraQ} restricted by \eqref{a1d4-extra} and
\eqref{a1a3-extra}.

\subsubsection{$(A_1,D_4)$ 
partition function}

We first study the $(A_1,D_4)$ partition function. Using
\eqref{a1d4paraQ} restricted by \eqref{a1d4-extra}, we see that the coefficients
$D_i$ in the strong coupling expansion \eqref{eq:final-D4} are
rewritten as
\begin{align}
D_1 &= \Big(3\nu^3 - 2\nu(\theta_s^2-\theta_s\theta_t+\theta_t^2) -
 \frac{2}{9}(2\theta_s-\theta_t)(\theta_s+\theta_t)(\theta_s-2\theta_t)\Big)
 -\frac{3\nu}{4}Q^2~,
\\
\nonumber
D_2 &=\frac{9 \nu ^6}{2} +
\nu ^4 \Big(\frac{105}{16}-6 (\theta_s^2-\theta_s \theta_t+\theta_t^2)\Big)-\frac{2}{3} \nu ^3 (\theta_s-2 \theta_t) (2 \theta_s-\theta_t) (\theta_s+\theta_t)
\\\nonumber
&\qquad +\nu ^2 \Big(\frac{1}{2} (\theta_s^2-\theta_s \theta _t+\theta _t^2) (4 (\theta_s^2-\theta_s \theta_t+\theta _t^2)-11)+\frac{3}{16}\Big)
\\\nonumber
&\qquad +\frac{\nu}{9}  (2 \theta_s-\theta_t) (\theta_s+\theta_t) (\theta_s-2 \theta_t) (4 (\theta_s^2-\theta_s \theta_t+\theta _t^2)-7)
\\\nonumber
&\qquad +\frac{1}{324} \Big(32 \theta_s^6-96 \theta_s^5 \theta _t+\theta_s^4 (36-24 \theta_t^2)
+8 \theta_s^3 \theta_t (26 \theta_t^2-9)-3 \theta_s^2 (8 \theta _t^4-36 \theta_t^2+9)
\\\nonumber
&\qquad \qquad -3 \theta_s \theta_t (32 \theta _t^4+24 \theta_t^2-9)+\theta_t^2 (32 \theta_t^4+36 \theta_t^2-27)\Big)\\\nonumber
&\qquad + \frac{1}{192}\Big(-432\nu^4 + 18\nu^2(16(\theta_s^2-\theta_s\theta_t+\theta_t^2)-37) +32\nu(2\theta_s-\theta_t)(\theta_s+\theta_t)(\theta_s-2\theta_t)\\
&\qquad \qquad + 104(\theta_s^2-\theta_s\theta_t+\theta_t^2) - 3 \Big)Q^2 + \frac{1}{256}(72\nu^2+25)Q^4~.
\end{align}
Note that they are invariant under the action of $S_3$ generated by
\eqref{a1d4Weyl1} and \eqref{a1d4Weyl2}, as expected. Moreover, they are also
invariant under $Q\to -Q$, or equivalently
$(\epsilon_1,\epsilon_2)\to (-\epsilon_1,-\epsilon_2)$. We have checked
this invariance up to $D_4$. We will briefly comment on 
this symmetry in the next section.

\subsubsection{$(A_1,A_3)$ 
partition function}

Let us turn to the $(A_1,A_3)$ partition function.
Using \eqref{a1a3paraQ} restricted by \eqref{a1a3-extra}, the
coefficients $D_k$ in the strong coupling expansion \eqref{eq:final-A3}
are rewritten as
\begin{align}
D_1 &= -\frac{i\nu(68\nu^2-9\theta^2+2)}{36}  + \frac{19}{36}i\nu Q^2~,
\\[1mm]
D_2 &= -\frac{289}{162}\nu^6 + \frac{153\theta^2-1159}{324}\nu^4-\Big(\frac{\theta^4}{32}-\frac{11\theta^2}{18}+\frac{271}{648}\Big)\nu^2-\frac{\theta^2(11\theta^2-68)}{1728}\\
&\quad + \Bigg(\frac{323}{324}\nu^4 -\Big(
 \frac{19\theta^2}{144}-\frac{349}{162}\Big)\nu^2 -
 \frac{71\theta^2}{864} + \frac{17}{432} \Bigg)Q^2 -
 \Big(\frac{361}{2592}\nu^2+\frac{131}{1728}\Big) Q^4 ~,
\nonumber
\\[2mm]
D_3 &= \frac{4913 i
}{4374}\nu^9 -\frac{17 i \left(153 \theta ^2-2284\right)}{5832}\nu^7+\frac{i
 \left(1377 \theta ^4-47178 \theta ^2+279464\right)}{23328}\nu^5
\nonumber\\[2mm]
& \quad -\frac{i \left(9 \theta ^2 \left(81 \theta ^4-5072 \theta
 ^2+77876\right)-899576\right)}{279936}\nu^3
-\frac{i \left(99 \theta
 ^6-4270 \theta ^4+28504 \theta ^2-3360\right)}{62208}\nu
\nonumber\\
&\quad +\Bigg(-\frac{5491 i \nu ^7}{5832}+\frac{i \left(2907 \theta ^2-68839\right)}{11664}\nu^5-\frac{i
 \left(1539 \theta ^4-94152 \theta ^2+1161988\right)}{93312}\nu^3
\nonumber\\
&\quad -\frac{i \left(1487 \theta ^4-61300 \theta ^2+66264\right)
 }{62208}\nu
\Bigg)Q^2 + \Bigg(\frac{6137 i}{23328}\nu^5  -\frac{i \left(3249 \theta
 ^2-118736\right)}{93312}\nu^3
\nonumber\\
&\quad  -\frac{i \left(3877 \theta ^2-92390\right) \nu }{62208} \Bigg)Q^4 +
 \Bigg( -\frac{6859 i \nu ^3}{279936}-\frac{2489 i \nu }{62208} \Bigg)Q^6~.
\end{align}
These coefficients are invariant under the $S_2$-action
\eqref{a1a3Weyl}, as expected. Moreover, we see that they 
are also invariant under $Q\to -Q$, or
equivalently $(\epsilon_1,\epsilon_2)\to (-\epsilon_1,-\epsilon_2)$.  We
will briefly comment on 
this symmetry in the next section.


\section{Conclusions and discussions}
\label{sec:conclusion}

In this paper, we have evaluated Liouville irregular states of rank-two
and three, based on an ansatz proposed in \cite{Gaiotto:2012sf}.
Using
these two irregular states, we have computed Liouville irregular
conformal blocks corresponding to the partition functions of the $(A_1,A_3)$ and
$(A_1,D_4)$ theories for general $\Omega$-background parameters
$\epsilon_1$ and $\epsilon_2$. In the limit $\epsilon_1+\epsilon_2\to
0$, our result correctly reproduces the strong coupling expansions of
the partition functions obtained from 
the tau-functions of Painlev\'e II and IV. This confirms that
the ansatz proposed in \cite{Gaiotto:2012sf} gives consistent results with \cite{Nagoya:2015cja, Nagoya:2016mlj, Nagoya:2018}. Moreover, we have shown that the asymptotic
expansion of an irregular state proposed in \cite{Gaiotto:2012sf} corresponds to the strong coupling
expansion of partition functions studied in \cite{Bonelli:2016qwg}. In
addition, our
result on $(A_1,A_3)$ implies that the irregular conformal block with one irregular
singularity of rank $3$ is also related to
the tau-function of Painlev\'e
II, in a similar way as 
that
with one
irregular singularity of rank $\frac{3}{2}$ and a regular singularity \cite{Nagoya:2018}.

We have also shown that 
our partition functions 
are invariant under the
action of the Weyl group of the flavor symmetry when the Liouville
momentum of irregular states is appropriately identified with the
linear combination of a mass parameter and the $\Omega$-background
parameters. From this observation, we have
conjectured a general relation \eqref{eq:dictionary} between the
Liouville momentum of rank-$n$ irregular state and
the corresponding mass
parameter in four dimensions. We have also given an interpretation of
this conjectured relation in terms of the colliding limit of regular
singularities.

With the conjectured relation \eqref{eq:dictionary}, we find that the partition
functions $\mathcal{Z}_{(A_1,A_3)}^{\epsilon_1,\epsilon_2}$ and $\mathcal{Z}_{(A_1,D_4)}^{\epsilon_1,\epsilon_2}$ are
invariant under $(\epsilon_1,\epsilon_2)\to(-\epsilon_1,-\epsilon_2)$
with masses and the vev of Coulomb branch operators fixed. This
invariance is consistent with quantum periods of
Argyres-Douglas theories recently evaluated in \cite{Ito:2017iba,
Ito:2018hwp, Ito:2019twh}. The quantum periods are $a_I \equiv \oint_{A_I} \lambda$
and $a_{D\,I} \equiv \oint_{B_I} \lambda$ deformed by the
$\Omega$-background $(\epsilon_1,\epsilon_2) = (\hbar,0)$, where
$A_I$ and $B_I$ are canonical 1-cycles of
the Seiberg-Witten curve, and $\lambda$ is the $\Omega$-deformed Seiberg-Witten
1-form. These periods are related to the deformed prepotential
${\displaystyle \mathcal{F}\equiv \lim_{(\epsilon_1,\epsilon_2)\to
(\hbar,0)}(-\epsilon_1\epsilon_2\log\mathcal{Z}^{\epsilon_1,\epsilon_2})}$ by $a_D = \partial \mathcal{F}/\partial a$. As shown in
\cite{Ito:2018hwp, Ito:2019twh},  the $\hbar$-expansion of these quantum
periods for $(A_1,A_r)$ and
$(A_1,D_r)$ theories have only {\it even} powers
of $\hbar$.\footnote{In \cite{Ito:2018hwp}, the $(A_1,A_3)$ and
$(A_1,D_4)$ theories are
realized at the most singular point on the Coulomb branch of $\SU(2)$ gauge
theory with two and three flavors, respectively. Its generalization to
the whole $(A_1,A_r)$ and $(A_1,D_r)$ theories was carefully studied in \cite{Ito:2019twh}. See \cite{Ito:2017iba} for discussions around
the monopole point on the Coulomb branch.} This means that, at least in
the limit of $\epsilon_2\to 0$, the quantity $\epsilon_1\epsilon_2\log\mathcal{Z}^{\epsilon_1,\epsilon_2}$ is invariant
under $(\epsilon_1,\epsilon_2)\to (-\epsilon_1,-\epsilon_2)$. Our result
suggests that this invariance extends to the case of
$\epsilon_1,\epsilon_2\neq 0$. We leave a detailed study of this symmetry for future work.




\section*{Acknowledgements}

The authors thank Katsushi Ito, Yusuke Kimura, Kazunobu Maruyoshi, Takafumi Okubo, Makoto Sakaguchi, Sakura Sch\"afer-Nameki and Yuji Sugawara for various discussions. T.~N. particularly
thanks Matthew Buican for various illuminating dicussions in
many collaborations on related topics. The authors also thank the
organizers of the international workshop ``KEK Theory workshop 2018,''
where they had many useful discussions on this topic.
The work of T.~N. is partially supported by JSPS Grant-in-Aid for Early-Career Scientists 18K13547.
The work of T.~U. is supported by Grant-in-Aid for JSPS Research Fellows 19J11212.

\appendix


\section{Computations of generalized descendants}
\label{app:strategy}

In this appendix, we explain a little more about the computations of
generalized descendants involved in irregular states of rank-two and
three, following 
\cite{Gaiotto:2012sf}.

\subsection{Rank-two state  $|I^{(2)}(\alpha,\pmb{c};\pmb{\beta})\rangle$}
\label{app:detail-D4}

Let us start with the rank-two irregular state 
$|I^{(2)}(\alpha,\pmb{c};\pmb{\beta})\rangle$, where
$\pmb{c}=(c_1,c_2)$ and $\pmb{\beta}=(\beta_0,\beta_1)$. According to
\cite{Gaiotto:2012sf}, this state is expected to be expanded as
\eqref{eq:expansion-D4} in terms of generalized descendants
$|I^{(1)}_k\rangle$ for $k=1,2,\cdots$. These states are fixed
order by order so that \eqref{eq:expansion-D4} satisfies
\eqref{eq:L4-D4} -- \eqref{eq:L0-D4}.
%
%
%
Indeed, 
substituting 
\eqref{eq:expansion-D4} in \eqref{eq:L4-D4} -- \eqref{eq:L0-D4} leads to
the following equations for each $k$:
\begin{align}
L_{n>4}|I_{k}^{(1)}\rangle &= 0~,
\label{eq:L5-2k}
\\
L_4|I_{k}^{(1)}\rangle &= -|I_{k-2}^{(1)}\rangle~,
\\
L_3|I_{k}^{(1)}\rangle &= -2c_1|I_{k-1}^{(1)}\rangle~,
\\
(L_2 + c_1^2)|I_{k}^{(1)}\rangle &= -(2\alpha-3Q)|I_{k-1}^{(1)}\rangle~,
\\
(L_1+2c_1(\beta-Q))|I_{k}^{(1)}\rangle &=
 \left(\frac{\partial}{\partial c_1} +
 \frac{\nu_1}{c_1}\right)|I_{k-1}^{(1)}\rangle~,
\label{eq:L1-2k}
\\
L_0|I_{k}^{(1)}\rangle &= \left(\Delta_{\beta} + 2k +
 c_1\frac{\partial}{\partial c_1}\right)|I_{k}^{(1)}\rangle~.
\end{align}
Solving these equations successively for $k=1,2,\cdots$, we obtain
$|I^{(1)}_k\rangle$ for all $k$.
To be concrete, in evaluating $|I^{(1)}_k\rangle$,
we write down the most general linear combination of level-$2k$
generalized desendants, and then fix 
their coefficients by demanding that the equations \eqref{eq:L5-2k} --
\eqref{eq:L1-2k} are satisfied. Note that $L_{-1},\,1/c_1$ and
$\partial_{c_1}$ are all regarded as level-one.
The results for $k=1$ and $k=2$ are shown in \eqref{eq:descendant2} and
 \eqref{eq:descendant3}.
In appendix \ref{app:D3 with Q}, we show the coefficients $D_k$
 appearing in \eqref{eq:result-D4} for $k=1,\cdots,4$. For computing
 $D_k$ for $k=1,\cdots,4$, we need
 to evaluate  $|I^{(1)}_k\rangle$ for $k=1,\cdots,5$.

\subsection{
Rank-three state $|I^{(3)}(\alpha,\pmb{c};\pmb{\beta})\rangle$}
\label{app:A3-detail}

Let us turn to the rank-three irregular state
$|I^{(3)}(\alpha,\pmb{c};\pmb{\beta})\rangle$, where
$\pmb{c}=(c_1,c_2,c_3)$ and $\pmb{\beta}=(\beta_0,\beta_1,\beta_2)$.
%
%
%
%
As in Sec.~\ref{app:detail-D4}, we fix the generalized descendants
$|I^{(2)}_k\rangle$ in the ansatz \eqref{eq:expansion-A3} by demanding
that \eqref{eq:expansion-A3} satisfies \eqref{eq:L6-A3} --
\eqref{eq:L0-A3}.
To that end, we substitute
\eqref{eq:expansion-A3} in \eqref{eq:L6-A3} -- \eqref{eq:L0-A3}
to obtain
\begin{align}
L_{n>6}|I^{(2)}_{k}\rangle &=0~,
\\
L_6|I^{(2)}_{k}\rangle &= -|I^{(2)}_{k-2}\rangle
\\
L_5|I^{(2)}_{k}\rangle &= -2c_2|I^{(2)}_{k-2}\rangle~,
\\
(L_4 +c_2^2)|I^{(2)}_{k}\rangle &= -2c_1|I^{(2)}_{k-1}\rangle~,
\\
(L_3+2c_1c_2)|I^{(2)}_{k}\rangle &= - 2(\alpha-2Q)|I^{(2)}_{k-1}\rangle~,
\\
(L_2 + c_2(2\beta-3Q)+c_1^2)|I^{(2)}_{k}\rangle &=  \left(\partial_{c_1} -\frac{2c_1}{c_2}(\alpha-\beta) \right)|I^{(2)}_{k-1}\rangle~,
\\
(L_1 +2c_1(\beta-Q)-c_2\partial_{c_1})|I^{(2)}_{k}\rangle &= \left(2\partial_{c_2}+ \frac{2c_1^2}{c_2^2}(\alpha-\beta) + \frac{2\rho_2}{c_2} \right)|I^{(2)}_{k-1}\rangle~,
\\
 L_0|I^{(2)}_{k}\rangle &= \left(\Delta_\beta +3k+c_1\partial_{c_1} + 2c_2 \partial_{c_2}\right)|I^{(2)}_{k}\rangle~.
\end{align}
We solve these equations successively for $k=1,2,\cdots$. The strategy
is again to write down the most general linear combination of
level-$3k$ generalized descendants for $|I^{(2)}_k\rangle$, and fix
their coefficients so that the above equations are satisfied.
Note here that $L_{-1},\,1/c_1$ and $\partial_{c_1}$ are regarded as
level-one, and $1/c_2$ and $\partial_{c_2}$ are regarded as level-two.

As discussed 
in the \ref{sec:A1A3}, we only need to evaluate the leading
contributions in the $c_1\to 0$ limit. Below, we write down
expressions for the first several $|I^{(2)}_k\rangle$:
\begin{align}
 |I^{(2)}_0\rangle &= |I^{(2)}(\beta,c_1,c_2)\rangle~,
\\[1mm]
|I^{(2)}_1\rangle &= \left(\frac{1}{3c_2}L_{-1} + \frac{3\alpha-
 4\beta}{3c_2}\partial_{c_1} -\frac{2c_1}{3c_2}\partial_{c_2}\right.
\nonumber\\
&\qquad \qquad  - \left.\frac{2 \left(\rho_2-\beta ^2+\beta  \alpha+Q (\beta -\alpha )\right)c_1}{c_2^2}+\frac{2 (\beta -\alpha )c_1^3}{3c_2^3}
\right)|I^{(2)}(\beta,c_1,c_2)\rangle~,
\\[2mm]
 |I^{(2)}_2\rangle  &= \Bigg(\frac{\rho_4}{c_2^3} - \frac{1}{6c_2^2}L_{-2} + \frac{\frac{Q}{6}+\frac{19}{9}\beta -\frac{5\alpha}{3}}{c_2^2}\partial_{c_2} +\frac{1}{18c_2^2}L_{-1}^2 + \frac{(-4\beta+3\alpha)}{9c_2^2}L_{-1}\partial_{c_1}
\nonumber\\
&\qquad \qquad + \frac{-3+64\beta^2-96\beta\alpha + 36\alpha^2}{72c_2^2}\partial_{c_1}^2+\mathcal{O}(c_1)\Bigg)|I^{(2)}(\beta,c_1,c_2)\rangle~,
\end{align}
 where $\rho_4$ is determined as
 \begin{align}
  \rho_4= \frac{1}{12} \left(\beta _3-\alpha \right) \left(4 \alpha ^2-26 \alpha  \beta _3+34 \beta _3^2+35 Q^2+27 \alpha  Q-69 \beta _3 Q-1\right)~.
 \end{align}
In Sec.~\ref{sec:A1A3}, we write down expressions for $D_1,\cdots,D_3$
appearing in the expansion \eqref{eq:pre-final-A3}. To evaluate these three
coefficients, we need to evaluate $|I^{(2)}_k\rangle$ for
$k=1,\cdots,6$, up to sub-leading terms in the limit $c_1\to 0$.

\section{More coefficients for the $(A_1,D_4)$ partition function
}
\label{app:D3 with Q}

We here show the coefficients $D_i$ in \eqref{eq:Q=0-D4} up to $i=4$
using \eqref{a1d4paraQ} and \eqref{a1d4-extra}. 
These
coefficients are 
written in terms of
symmetric polynomials in the variables $X_1,X_2$ and $X_3$, where
\begin{align}
\begin{aligned}
&X_1=\theta_s-2\theta_t~,\\
&X_2=\theta_t-2\theta_s~,\\
&X_3=\theta_s+\theta_t~.
\end{aligned}
\end{align}
This reflects the fact that the mass parameters are permuted by the
action of the Weyl group of the flavor $\SU(3)$ symmetry
of the $(A_1,D_4)$ theory, as discussed in Sec.~\ref{sec:flavor}.
Indeed, these three variables are 
permuted under the $S_3$ Weyl
transformations \eqref{a1d4Weyl1} and  \eqref{a1d4Weyl2}. 
The elementary symmetric polynomials $s_1,s_2$ and $s_3$ of these
variables are given by
\begin{align}
\begin{aligned}
&s_1=X_1+X_2+X_3=0~,\\
&s_2=X_1X_2+X_2X_3+X_3X_1=-3(\theta_s^2-\theta_s\theta_t+\theta_t^2),\\
&s_3=X_1X_2X_3=(\theta_s-2\theta_t)(\theta_t-2\theta_s)(\theta_s+\theta_t)~.
\end{aligned}
\label{eq:s1s2s3}
\end{align}
Since the partition function is supposed to be invariant under the
action of the Weyl
group, the coefficients $D_i$ of its strong coupling expansion are expected to be written in terms of $s_2$ and $s_3$.
Indeed, using \eqref{eq:s1s2s3}, 
we may rewrite $D_1$ and $D_2$ as 
\begin{align}
D_1&=3 \nu ^3+\frac{2\nu s_2 }{3}+\frac{2 s_3}{9} -\frac{3 \nu}{4}Q^2~,
\\\nonumber
D_2&= \frac{9 \nu ^6}{2}+\nu ^4 \left(2 s_2+\frac{105}{16}\right) +\frac{2 \nu ^3 s_3}{3} +\frac{\nu ^2}{144} (8 s_2 (4 s_2+33)+27)+\frac{\nu}{27}  (4 s_2+21) s_3
\\\nonumber
&\quad\qquad +\frac{1}{324} (s_2 (4 s_2+9)+8 s_3^2)
\\\nonumber
&\quad +\Bigg(-\frac{9 \nu ^4}{4}+\nu ^2 \left(-\frac{s_2}{2}-\frac{111}{32}\right)-\frac{\nu  s_3}{6}+\left(-\frac{13 s_2}{72}-\frac{1}{64}\right)\Bigg)Q^2
\\
&\quad +\left(\frac{9 \nu ^2}{32}+\frac{25}{256}\right)Q^4~.
\end{align}
Moreover, $D_3$ and $D_4$ are also evaluated as
\begin{align}
\nonumber
D_3&= -\frac{9 \nu ^9}{2}+\nu ^7 \left(-3 s_2-\frac{315}{16}\right)-\nu ^6 s_3+\nu ^5 \left(-\frac{2 s_2^2}{3}-\frac{79 s_2}{8}-\frac{411}{16}\right)-\frac{\nu ^4}{72} (32 s_2+273) s_3
\\\nonumber
&\quad\qquad +\frac{\nu ^3}{648} \left(-32 s_2^3-816 s_2^2-5535 s_2-48 s_3^2-1701\right)-\frac{\nu ^2}{648} \left(32 s_2^2+600 s_2+2475\right) s_3
\\\nonumber
&\quad\qquad +\frac{\nu}{486}  \left(-4 s_2^3-105 s_2^2-s_2 \left(8 s_3^2+297\right)-84 s_3^2\right)-\frac{1}{4374}s_3 \left(12 s_2^2+351 s_2+8 s_3^2+972\right)
\\\nonumber
&\quad +\Bigg(\frac{27 \nu ^7}{8}+\nu ^5 \left(\frac{3 s_2}{2}+\frac{981}{64}\right)+\frac{\nu ^4 s_3}{2}+\frac{1}{48} \nu ^3 \left(8 s_2^2+203 s_2+1134\right)+\nu ^2 \left(\frac{s_2}{9}+\frac{65}{48}\right)
\\\nonumber
&\quad\qquad +\frac{1}{864} \nu  \left(112 s_2^2+2307 s_2+16 s_3^2+621\right)) s_3+\frac{1}{2592}(104 s_2+1305) s_3\Bigg)Q^2
\\\nonumber
&\quad +\left(-\frac{27 \nu ^5}{32}-\frac{3}{256} \nu ^3 (16 s_2+247)-\frac{\nu ^2 s_3}{16}+\frac{\nu}{768} (-154 s_2-1695)-\frac{25 s_3}{1152}\right)Q^4
\\
&\quad +\left(\frac{9 \nu ^3}{128}+\frac{75 \nu }{1024}\right) Q^6~,
\\\nonumber
D_4&= -\frac{27 \nu ^{12}}{8}+\nu ^{10}\left(-3 s_2-\frac{945}{32}\right) -\nu ^9s_3 +\nu ^8\left(-s_2^2-\frac{171 s_2}{8}-\frac{50049}{512}\right) -\frac{\nu ^7}{24} (16 s_2+189) s_3
\\\nonumber
&\quad\qquad +\nu ^6\left(-\frac{4 s_2^3}{27}-\frac{373 s_2^2}{72}-\frac{1737 s_2}{32}-\frac{s_3^2}{9}-\frac{34335}{256}\right)-\frac{\nu ^5}{432} \left(64 s_2^2+1620 s_2+9567\right) s_3 
\\\nonumber
&\quad\qquad +\nu ^4\left(-\frac{2 s_2^4}{243}-\frac{35 s_2^3}{81}-\frac{1153 s_2^2}{144}+\left(-\frac{4 s_3^2}{81}-\frac{3323}{64}\right) s_2-\frac{49 s_3^2}{72}-\frac{15381}{512}\right) 
\\\nonumber
&\quad\qquad -\frac{\nu ^3}{11664}s_3 \left(128 s_2^3+5280 s_2^2+70740 s_2+64 s_3^2+273861\right) 
\\\nonumber
&\quad\qquad +\frac{\nu ^2}{93312}\Big(-256 s_2^4-14976 s_2^3-8 \left(64 s_3^2+30969\right) s_2^2-18 \left(832 s_3^2+45063\right) s_2
\\\nonumber
&\quad\qquad\qquad -9 \left(11888 s_3^2+5103\right)\Big) 
\\\nonumber
&\quad\qquad -\frac{\nu }{52488}s_3 \left(96 s_2^3+5616 s_2^2+64 s_3^2 s_2+77274 s_2+1008 s_3^2+207765\right)
\\\nonumber
&\quad\qquad +\frac{1}{629856}\Big(-48 s_2^4-2808 s_2^3-3 \left(64 s_3^2+8829\right) s_2^2-27 \left(400 s_3^2+1701\right) s_2
\\\nonumber
&\quad\qquad\qquad -16 s_3^2 \left(4 s_3^2+4617\right)\Big)
\\\nonumber
&\quad +\Bigg(\frac{27 \nu ^{10}}{8}+\frac{9\nu ^8}{8} (2 s_2+27) +\frac{3 s_3 \nu ^7}{4}+\frac{\nu ^6}{512} \left(256 s_2^2+7760 s_2+57555\right) +\frac{\nu ^5}{96} (32 s_2+495) s_3 
\\\nonumber
&\quad\qquad +\nu ^4\left(\frac{s_2^3}{27}+\frac{299 s_2^2}{144}+\frac{4803 s_2}{128}+\frac{s_3^2}{18}+\frac{193455}{1024}\right)+\frac{\nu ^3}{864} \left(32 s_2^2+1148 s_2+10611\right) s_3
\\\nonumber
&\quad\qquad +\frac{\nu ^2}{82944}\left(3840 s_2^3+190656 s_2^2+8 \left(128 s_3^2+338931\right) s_2+9 \left(1984 s_3^2+168183\right)\right) 
\\\nonumber
&\quad\qquad +\frac{\nu }{46656}s_3 \left(1344 s_2^2+52380 s_2+64 s_3^2+426951\right)
\\\nonumber
&\quad\qquad +\frac{1}{373248}(832 s_2^3+85752 s_2^2+1664 s_3^2 s_2+387018 s_2+41616 s_3^2+15309)\Bigg) Q^2
\\\nonumber
&\quad +\Bigg(-\frac{81 \nu ^8}{64}-\frac{9\nu ^6}{256} (16 s_2+287) -\frac{3\nu ^5 s_3 }{16}+\frac{\nu ^4}{4096}\left(-256 s_2^2-11680 s_2-126603\right) 
\\\nonumber
&\quad\qquad -\frac{\nu ^3}{384} (16 s_2+331) s_3 +\frac{\nu ^2}{36864}\left(-4256 s_2^2-157176 s_2-256 s_3^2-1416717\right)
\\\nonumber
&\quad\qquad -\frac{\nu }{6912}(308 s_2+6507) s_3 +\frac{1}{663552}(-11616 s_2^2-873720 s_2-1600 s_3^2-378189)\Bigg) Q^4
\\\nonumber
&\quad +\Bigg(\frac{27 \nu ^6}{128}+\frac{3\nu ^4}{128} (2 s_2+51) +\frac{\nu ^3s_3 }{64}+\frac{3\nu ^2}{8192} (272 s_2+5433) 
\\\nonumber
&\quad\qquad +\frac{25\nu s_3}{1536} +\frac{1}{147456}(2600 s_2+98613)\Bigg) Q^6
\\
&\quad +\left(-\frac{27 \nu ^4}{2048}-\frac{225 \nu ^2}{8192}-\frac{625}{131072}\right) Q^8~.
\end{align}

\bibliography{irreg1}
\bibliographystyle{utphys}

\end{document}